\documentclass[usenatbib]{mn2e}
\usepackage{epsfig}
\usepackage{threeparttable}
\usepackage{footnote}
\usepackage{cite}
\usepackage{booktabs}
\usepackage{url}
\usepackage{caption}
\usepackage{subfig}
\usepackage{hyperref}
\usepackage{graphicx}

\begin{document}


\title[An IFU view of HCG 31 ]
{Probing the nature of the pre-merging system\\ Hickson Compact Group 31 through IFU data. 
}

\author[M. Alfaro-Cuello et al.]
{
\parbox[t]{\textwidth}{
 M. Alfaro-Cuello$^1$, S. Torres-Flores$^1$, E. R. Carrasco$^2$, C. Mendes de Oliveira$^3$, D. F. de Mello$^4$ \& P. Amram$^5$ 
 }
\vspace*{6pt} \\
$^1$ Departamento de F\'isica y Astronom\'ia, Universidad de La Serena, Av. Cisternas 1200, La Serena, Chile.\\
$^2$ Gemini Observatory/AURA, Southern Operations Center, Casilla 603, La Serena, Chile. \\
$^3$ Instituto de Astronomia, Geof\'isica e Ci\^encias Atmosf\'ericas da Universidade de S\~{a}o Paulo, Cidade Universit\'aria, CEP:05508-900,\\
S\~{a}o Paulo, SP, Brazil.\\
$^4$ Catholic University of America, Washington, DC 20064, USA.\\
$^5$ Aix Marseille Universit\'e, CNRS, Laboratoire d'Astrophysique de Marseille, France.\\
}

\date{\today}

\pagerange{\pageref{firstpage}--\pageref{lastpage}}

\maketitle

\label{firstpage}

\begin{abstract}
We present a study of the kinematics and the physical properties of the central region of the Hickson Compact Group 31 
(HCG 31), focusing on the HCG 31A+C system, using integral field spectroscopy data taken with the Gemini-south telescope. The main 
players in the merging event (galaxies A and C) are two dwarf galaxies that in the past have already had one close encounter, 
given the observed tidal tails, and may now be in their second approach, possibly about to merge. 
We present new velocity fields and H$\alpha$ emission, stellar continuum, velocity dispersion, electron density, H$\alpha$ 
equivalent-width and age maps. Considering the high spatial resolution of the IFU data, we were able to measure various 
components and estimate their physical parameters, spatially resolving the different structures in this region. 
Our main findings are the following: (1) we report for the first time the presence of a super stellar cluster next to the
burst associated to the HCG 31C central blob, related to the high values of velocity dispersion observed in this region as well
as to the highest value of stellar continuum emission. This may suggest that this system is cleaning its environment through
strong stellar winds that may then trigger a strong star formation event in its neighborhood, 
(2) among other physical parameters, we estimate an \mbox{L(H$\alpha$)$\sim$14$\times$10$^{41}$\,erg\,s$^{-1}$} and a 
\mbox{SFR$\sim$11\,M$_{\odot}$\,yr$^{-1}$} for the central merging region of HCG 31 A+C. These values indicate a high star formation 
density, suggesting that the system is part of a merging object, supporting previous 
scenarios proposed for this system.
\end{abstract}


\thanks{Based on observations obtained at the Gemini Observatory, which is operated by the Association of Universities for 
Research in Astronomy, Inc., under a cooperative agreement with the NSF on behalf of the Gemini partnership: the National
Science Foundation (United States), the Science and Technology Facilities Council (United Kingdom), the National Research 
Council (Canada), CONICYT (Chile), the Australian Research Council (Australia), Minist\'erio da Ci\^encia e Tecnologia (Brazil)
and Ministerio de Ciencia, Tecnolog\'ia e Innovaci\'on Productiva (Argentina) -- Observing runs: GS-2012B-Q-60.}

\section{Introduction}

Local compact groups of galaxies are excellent places where to study galaxy transformation. As a consequence of the strong
gravitational encounters, galaxies in compact groups develop tidal tails, bursts of star formation, gas flows, nuclear activity
and kinematic perturbations. All these phenomena have strong influences on the evolution of galaxies in compact groups. For
example, simulations suggest the existence of gas inflows in merging systems. These flows can mix the metal content of the 
interacting galaxies (Rupke et al. 2010a, Perez et al. 2011), and can enhance the star formation of the central parts of the
interacting/merging systems, increasing their IR luminosities. On the other hand, such interactions can affect the global 
kinematics of the galaxies in the compact group. This effect can be easily seen in the 2D velocity fields of such systems 
(Amram et al. 2003, Plana et al. 2003). Given that all the phenomena listed above are connected, it will be desirable to 
analyze all of them simultaneously, which in practice is not an easy task, given the lack of proper data and/or instrumental 
limitations. A way to circumvent some of these limitations is to use IFU (Integral Field Unit) Spectroscopic data, given that
this technique allows us to observe simultaneously the spectra of several regions, where the galaxy configuration is compact 
enough. We have obtained IFU observations of one of the most interesting compact groups of galaxies: Hickson Compact Group 31
(HCG 31, Hickson 1982), in order to study the evolutionary phase of this group and to search for a link between its kinematics
and chemical enrichment history. In addition, this system is an ideal laboratory to study star formation at different levels
(from small to giant star-forming regions), therefore, IFU data on HCG 31 will be extremely useful to understand the star
formation process in interacting/merging galaxies. 

The Hickson Compact Group 31 is an intriguing object because of its wide range of indicators of galaxy interaction and merging,
like multiple tidal tails, irregular morphology, complex kinematics, strong starburst spots and possible formation of tidal
dwarf galaxies (TDGs) (Rubin et al. 1990, L\'opez-S\'anchez et al. 2004, Mendes de Oliveira et al. 2006, Amram et al. 2007). 
HCG 31 is formed by several members: A+C, B, E, F, G, H, Q and R (Mendes de Oliveira et al.  2006). Galaxies A and C form the
main central part. These are late-type, gas-rich, large Magellanic-type irregular galaxies (Amram et al. 2007), which could be
responsible for the two optical tidal tails that extend towards the northeast and southwest of the A+C system. In addition, 
TDG candidates seem to have been formed out of  gas-rich material stripped from this system (L\'opez-S\'anchez et al. 2004). 
Two scenarios have been proposed to explain the nature of the system A+C: The system A+C is a single entity, i.e. a single 
interacting galaxy (Richer et al. 2003) and a second scenario supports the idea that there are two systems in a merging process
(Rubin et al. 1990). This latter scenario is supported by Amram et al. (2007)  who found that the dynamics of the A+C system 
has two main different components of approximately the same intensities, indicating that it is in a pre-merging stage.

In this paper we present integral field spectroscopic observations of the central region of HCG 31. We describe the Gemini
Integral Field Spectroscopic observations, data reductions and analysis in sections \S2 and \S3 respectively. Our results are
presented in section \S 4. The discussion and conclusions are presented in sections \S5 and \S6, respectively. We have adopted 
a  distance of 57.6\,Mpc assuming a redshift z=$0.0137$ (Hickson et al. 1992) and H$_0=72$\,km\,s$^{-1}$\,Mpc$^{-1}$.
       
\begin{figure*}
\hspace{1.5cm}
\includegraphics[width=\textwidth]{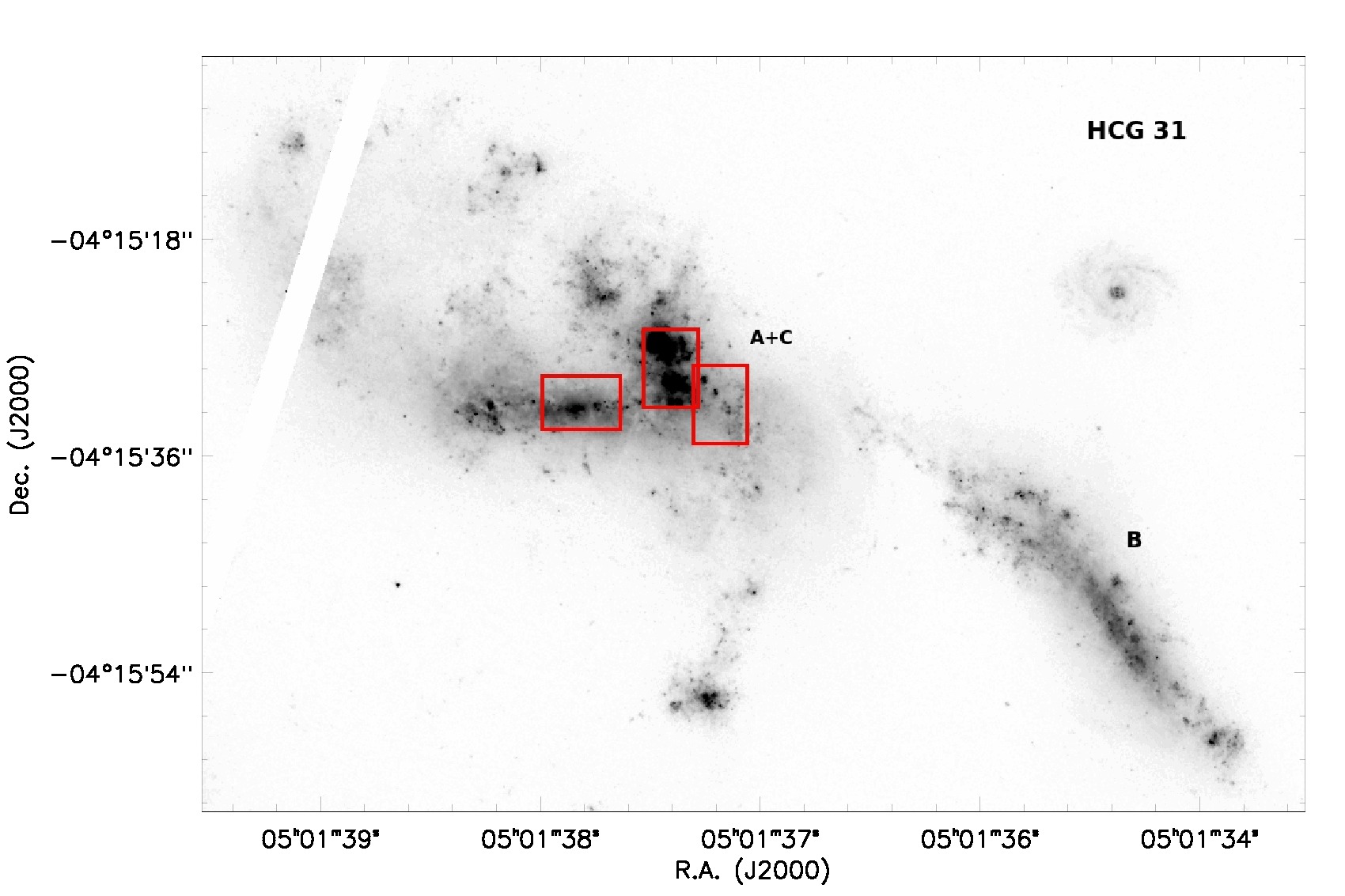}
\caption{\textit{Hubble Space Telescope} (HST) optical image (ACS, F435W) of HCG 31. The image shows the regions A+C, B and 
part of the southeast tidal tail. The red squares show the location of the three IFUs in the region A+C. These regions are
denominated, from left to right: region east, center and west. Image published by Torres-Flores et al. (2015).}
\label{hcg31}
\end{figure*}

\section{Observations and data reduction}

The central region of HCG 31 was observed with the Gemini South telescope during the nights between January 12 - 19, 2013,
with Gemini Multi-Object Spectrograph (GMOS, Hook et al. 2004) and the Integral Field Unit (IFU, Allington-Smith et al. 2002),
under the program GS-2012B-Q-60 (PI: S. Torres-Flores). In order to have a larger spatial coverage and map the whole A+C system,
three different fields were observed with the IFU in two-slit mode. The spectra and calibrations were acquired using the 
grating R400$+$r'-filter centered at 6300\,{\AA}. In the two-slit mode, the IFU covers a field of view of 7''$\times$5''. Each
IFU hexagonal element (the two-slit mode uses 1500 elements) has a projected scale of 0.2'' on the sky.  For each field, three
exposures of 700 sec each were obtained (no spectral or spatial dithering was applied during the observations). All fields were
observed during dark time (sky background 50\%) at an airmass between 1.23 and 1.31, with median seeing values of $\sim$0.5''
and 0.6'' and were not taken under photometric conditions (cloud cover: 70\% cirrus). Figure \ref{hcg31} shows an HST
optical image (ACS, F435W) with the location of the three GMOS/IFU fields superimposed onto it (``east'', ``center'' and 
``west'' fields, respectively). In addition, the spectrophotometric standard star LTT 4364 was observed in January 15, 2013 to
flux calibrate  the science spectrum. 

The observations were processed using a modified early version of the Gemini/GMOS IRAF package to reduce IFU
spectra\footnote{http://drforum.gemini.edu/topic/gmos-ifu-data-reduction-scripts/}. A master bias was generated by combined 
raw bias frames. Then all science exposures, comparison lamps, spectroscopic twilight flats and  Gemini Calibration Unit lamp 
flats (GCAL flats) were overscan-ed/trimmed, and bias subtracted using the master bias. Fibers identification was performed
with the task ``{\sc gfreduce}'' using high signal-to-noise GCAL flats. The resulting file with the fiber identifications was used 
later as a reference to identify the fibers  in the comparison lamps, science exposures and spectroscopic twilight flats. The
GCAL flats were processed by removing the GCAL$+$GMOS spectral response and the uneven illumination introduced by the GCAL unit.
The twilight flats were used to determine the relative fiber throughput and to correct for any  illumination pattern in the 
GCAL flat using the ``{\sc gfresponse}'' task. Wavelength calibration has been obtained using high signal-to-noise CuAr comparison
lamp using the ``{\sc gswavelength}'' routine. The residual values in the wavelength solution typically yielded a rms of 
$\sim$0.1\,{\AA}. The science exposure are then flat-fielded, corrected by the relative fiber throughput, wavelength calibrated,
sky subtracted and extracted. The same reduction steps were applied to process the standard star LTT 4364. The sensitivity
function was derived using the ``{\sc gsstandard}'' routine and the science exposures were then flux calibrated using the 
routine ``{\sc gscalibrate}''. 

The final step was the construction of data cubes for further analysis. Each processed 2D science image was transformed to a 3D 
data cube (x, y, $\lambda$) using the ``{\sc gfcube}'' routine. The output data cubes (3 per field) were re-sampled to square
pixels 0.1'' in size and corrected for atmospheric differential refraction. To produce a single data cube for each field,
the three cubes were combined using the program ``{\sc pymosaic}'' kindly provided by Dr. James Turner (2013, private 
communication). The final spectra have a resolution of $\sim$2.4{\AA} in FWHM (measured using the sky line at 6300\,{\AA}) and
a dispersion of $\sim$0.68{\AA}\ pix$^{-1}$, covering a wavelength interval of $\sim$5620{\AA}--6980{\AA}.

\section{Analysis}

\begin{figure*}
\includegraphics[width=1.5\columnwidth]{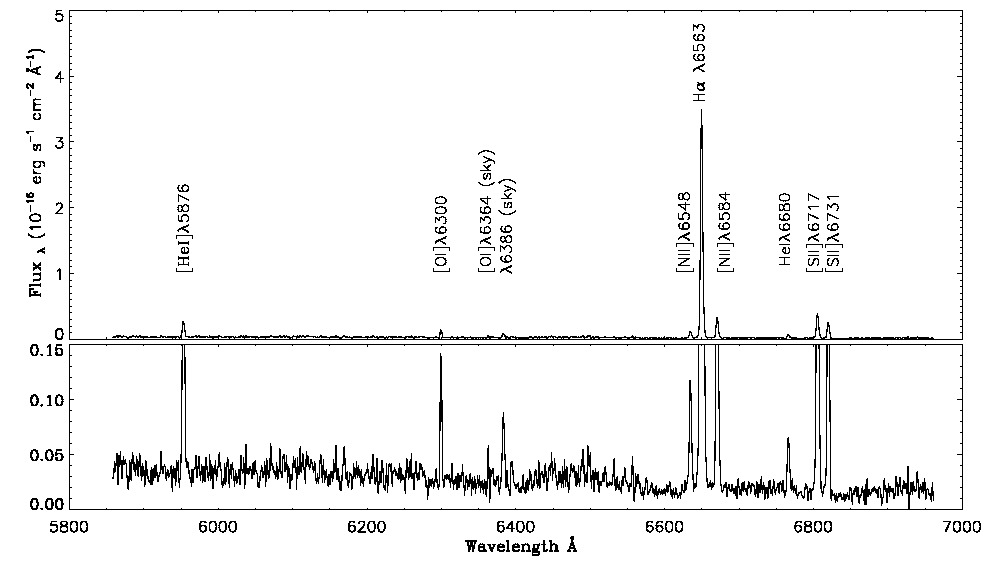}\\
\caption{Top spectrum: An example of the typical spectrum observed in the central region of HCG 31. The main emission lines are
labeled on this image. The position of the region which has this spectrum is shown in Fig. \ref{hcg31} (5''$\times$7'' centered
on RA: 05:01:37.77, DEC: -04:15:30.24). Bottom spectrum: Zoom in the continuum emission of the top spectrum.}
\label{espectro}
\end{figure*}

The optical emission of the central region of HCG 31A+C is dominated by two main bursts of star formation which coincide with 
the centers of galaxies HCG 31A and HCG 31C (Iglesias-P\'aramo \& V\'ilchez 1997). The ages of these bursts are 
very similar and therefore it is inferred that they were triggered by the interaction between galaxies A and C
(Iglesias-P\'aramo \& V\'ilchez 1997).
L\'opez-S\'anchez et al. (2004) detected some underlying stellar emission associated with these bursts. However, it 
is negligible compared with the strong emission coming from the nebular lines. In this work we will focus on the flux coming 
from the ionized gas. In Figure \ref{espectro} we show a typical spectrum observed in the central region of HCG 31A+C. This 
spectrum is dominated by strong emission lines, where the line ratios are typical of star-forming regions in the BPT plot
(Baldwin et al. 1981), with almost no emission coming from an underlying stellar population.

In Figure \ref{map_hst} we have included an optical image of the \textit{HST}/ACS (F435W) for the system 
HCG 31A+C, where we have over plotted the three IFU fields in red rectangles (5''$\times$7'' or $\sim$1355$\times$1897 pc each).
In addition, we have over plotted the position of the two main bursts of star formation detected by Iglesias-P\'aramo \& 
V\'ilchez et al. (1997) in yellow squares, which are located in the galaxies HCG 31C and HCG 31A (boxes have sizes 
1.1''$\times$1.1'', or $\sim$298$\times$298 pc). The box size was chosen considering the burst associated to the galaxy HCG 
31C, since this burst shows a stronger and more extended H$\alpha$ emission. The mean values of the different
physical parameters of these two main bursts were obtained considering this area.

\subsection{Extinction}

Observed flux should be affected by Galactic and internal extinctions. We have corrected the observed flux for Galactic 
extinction by using a color excess of \mbox{E(B-V)=0.04 mag} (NED database) and the Fitzpatrick (1999) extinction law.
The internal extinction clearly depends on the internal properties of the observed source. Starburst objects are usually dusty 
and, therefore, we expect to have high internal extinctions.  We note that the spectral coverage of our observations do not
allow us to derive the H$\alpha$/H$\beta$ ratio to determine the internal extinction through the Balmer decrement. Therefore,
in order to correct for this effect, we have used the spectra published by Mendes de Oliveira et al. (2006) to obtain an
average nebular color excess for the central region of HCG 31A+C. In this case, the nebular color excess was estimated by
using the observed H$\alpha$/H$\beta$ ratio and comparing it with the expected value listed in Osterbrock (1989), using the
recipes given in Dominguez et al. (2013). We obtain an average nebular color excess of E(B-V)$\sim$0.08\,mag. Given that
HCG 31A and HCG 31C are starburst objects (Gallagher et al. 2010), we have used the Calzetti et al. (2000) extinction law to 
correct our data by internal extinction, considering the nebular color excess estimated above (which is a factor of $\sim$2 
larger than the stellar color excess). We note that the extinction correction is not a critical issue for line ratios 
(given that the emission lines display a small wavelength separation). However, this correction is important for the estimation of
H$\alpha$ luminosities and star formation rates.

\subsection{Flux measurements}

The main nebular emission lines available in our data correspond to [NII]$\lambda$6548\,{\AA}, H$\alpha$$\lambda$6563\,{\AA},
[NII]$\lambda$6584\,{\AA}, [SII]$\lambda$6717\,{\AA} and [SII]$\lambda$6731\,{\AA}. The available spectral resolution of 
FWHM=2.4\,{\AA} ($\sim$114\,km\,s$^{-1}$, value adopted from method (ii), see next section) does not allow us to resolve
multiple components on each emission line profile. 
For this reason, fluxes have been estimated by fitting a single Gaussian on each observed line, by using {\sc fluxer}\footnote{ 
Interactive routine in {\sc idl} written by Christof Iserlohe. http://www.ciserlohe.de/fluxer/fluxer.html}
which allows us to determine the continuum level around the 
emission lines in a confident way. The stellar continuum definition was made in the wavelength range of 6620-6690\,{\AA}.
The output parameters of this procedure are flux, continuum, width, center, equivalent width and uncertainties derived from the
2D maps. Fluxes were compared with the values estimated by fitting a single Gaussian with the {\sc mpfit}\footnote{Fitting
program package in {\sc idl} written by Craig B. Markwardt, NASA/GSFC Code 662, Greenbelt, MD 20770.} code and with the task 
{\sc splot} in {\sc iraf}, and the results are fully consistent. Given that, we have included the H$\alpha$ emission
and stellar continuum emission maps in the top and bottom panels in Figure \ref{map_flux}, respectively.

\subsection{Velocity measurements}

The most intense emission line in our data set is H$\alpha$. This emission line is usually used to analyze the
kinematics of 
the warm gas in galaxies (e.g Amram et al. 2007). We used it to derive the velocity field of the central region of
HCG 31. As described above, the Gaussian fitting process gives us the center of each emission line. In this case, the estimated
center was converted into velocity, yielding a velocity field.
In a similar way, we used the width of the profiles to produce a velocity dispersion map. At first order, if we suppose that 
the different emission line profiles can be fitted by Gaussians, the velocity dispersion $\sigma$ corrected for instrumental 
and thermal broadening could be written as: $\sigma=\sqrt{\sigma_{obs}^2-\sigma_{inst}^2-\sigma_{th}^2}$, where $\sigma_{obs}$ 
is the measured velocity dispersion on the observed profile. 
The instrumental width $\sigma_{inst}$ was estimated in two different ways which resulted in consistent values: (i) from the 
average of the FWHM of the lines from the CuAr calibration lamp close to H$\alpha$, obtaining a spectral resolution of 
$\sim$2.3$\pm$0.1\,{\AA}, corresponding to $\sim$110\,km\,s$^{-1}$, leading to a resolution R$\sim$2850$\pm$125 (in H$\alpha$),
and (ii) following the method published by S\'anchez et al. (2012), who estimate the spectral resolution fitting a Gaussian
in the most intense sky lines in one frame. In our case, we have obtained an average of $\sim$2.4$\pm$0.1\,{\AA}, corresponding
to $\sim$114\,km\,s$^{-1}$, leading to a resolution R$\sim$2730$\pm$115. 
Finally, we adopt the value obtained from the second method. From this method we have estimated a instrumental width of
$\sigma_{inst}\sim$46$\pm$10\,km\,s$^{-1}$, where the uncertainty is represented by the standard deviation following S\'anchez
et al. (2012) method. We found a change in the value of the instrumental width across 
each observed fields. On average, this variation is of the order of $\sim$10\,km\,s$^{-1}$. However, we note that the
largest changes in this parameters occur at the edge of the fields, while in the centre of the fields the variation in the 
instrumental width is close to $\sim$6\,km\,s$^{-1}$. For the thermal width correction we have assumed a $\sigma_{th}$=9\,km\,s$^{-1}$, that 
corresponds to a electron temperature of 10000\,K (which is consistent with the temperatures estimated by L\'opez-S\'anchez et
al. 2004). The uncertainty associated with the corrected velocity dispersion was estimated by propagating in quadrature 
the uncertainties of $\sigma_{inst}$ and $\sigma_{obs}$.
On top and bottom panels in Figure \ref{map_vel}, we present the velocity and velocity dispersion maps, respectively.

\subsection{Star formation rates}

There are several tracers of star formation. For example, the UV radiation emerging from young hot stars can be used to estimate the star formation 
rate of galaxies. However, this emission considers stellar populations with ages up to $\sim$200 Myrs. Also, this emission is strongly affected by dust 
attenuation. To mitigate these problems, several authors use the UV and IR emission (e.g. Iglesias-P\'aramo et al. 2006 and more recently Eufrasio et al. 2014),
in order to take into account 
the emission of dust that is heated by the strong UV radiation field. Another way to estimate star formation rates is to use the H$\alpha$ emission, which
emerges from gas ionized by massive stars. Given the available data, the star formation rates in HCG 31 have been estimated by using its H$\alpha$ luminosity 
and the recipe given in Kennicutt (1998), which assumes a continuous star formation process, where 
\mbox{SFR$_{H\alpha}$=7.94$\times$10$^{-42}$$\times$L(H$_{\alpha}$)}.
However, we note that the SFRs derived from the H${\alpha}$ emission can be underestimated given that this emission is
affected by extinction, which corresponds to one of the most important sources of uncertainties in the determinations of SFR 
when using H${\alpha}$ luminosities (Kennicutt 1998). In the case of the current analysis, we have derived an average extinction
for the region under study, which comes from the Balmer decrement of archival longslit observations. Of course, the extinction
can vary along the extension of HCG 31 A+C, given the strong burst of star formation located in this region. Therefore, we 
caution the reader that our SFRs estimations can be affected by spatial variations in the extinction, therefore, our SFRs can
be overestimated (discussed in \S \ref{properties}).

\subsection{Electron densities}

Our spectroscopic data cover the [SII] lines, [SII]$\lambda$6717\,{\AA} and [SII]$\lambda$6731\,{\AA}. We have calculated the 
electron density (N$_e$) from the observed \mbox{[SII]$\lambda$6717\,{\AA}/[SII]$\lambda$6731\,{\AA}} ratio using the task 
{\sc temden}, which belongs to the {\sc iraf} {\sc stsdas} {\sc nebular} package. The task {\sc temden} implement the five-level
atomic model {\sc fivel} (De Robertis, Dufour \& Hunt 1987). 
We are not able to estimate the electron temperatures due to the spectral coverage ($\sim$5620\,{\AA}--6980\,{\AA}).
So, in this case we have assumed an electronic temperature of T$_e$=10000\,K. This assumption is
consistent with the electronic temperature found by L\'opez-S\'anchez et al. (2004) for the burst associated with HCG 31C
(T$_e$=10800$\pm$300\,K, Table 3 in that paper). The uncertainties in the estimates are represented by the standard deviation.
The electron density map obtained using this method is shown on the top panel of Figure \ref{map_abundance}.

\subsection{H$\alpha$ Equivalent Width and age determination}

We estimated the ages of the bursts by comparing the measured H$\alpha$ equivalent width (EW(H$\alpha$)) with Starburst99 
(Leitherer et al. 1999) modelled values. Considering the fact that the oxygen abundances in the regions of interest
range from 12+log(O/H)$\sim$8.2 and 12+log(O/H)$\sim$8.5 (L\'opez-S\'anchez et al. 2004,
Torres-Flores et al. 2015), the Starburst99 models were generated for a single stellar population with metallicities of Z=0.004
and Z=0.008\footnote{We used the conversion \mbox{Z=Z$_\odot \times$10$^{(log(O/H)-12+3.1)}$} and assumed
\mbox{log(O/H)$_\odot$=-3.10} (Fiorentino et al. 2012), and a solar metallicity of Z$_\odot$=0.018 (Allende-Prieto et
al. 2001).}, which correspond to Z=0.2--0.4Z$_\odot$, and Salpeter Initial Mass Function (IMF) with masses ranging between 
1--100 M$_\odot$. The map obtained for the estimated values of EW(H$\alpha$) is shown in the bottom panel of Figure 
\ref{map_abundance} and the age maps are shown in the top and bottom panels of Figure \ref{map_age}.

\begin{figure*}
\centering \includegraphics[width=2.\columnwidth]{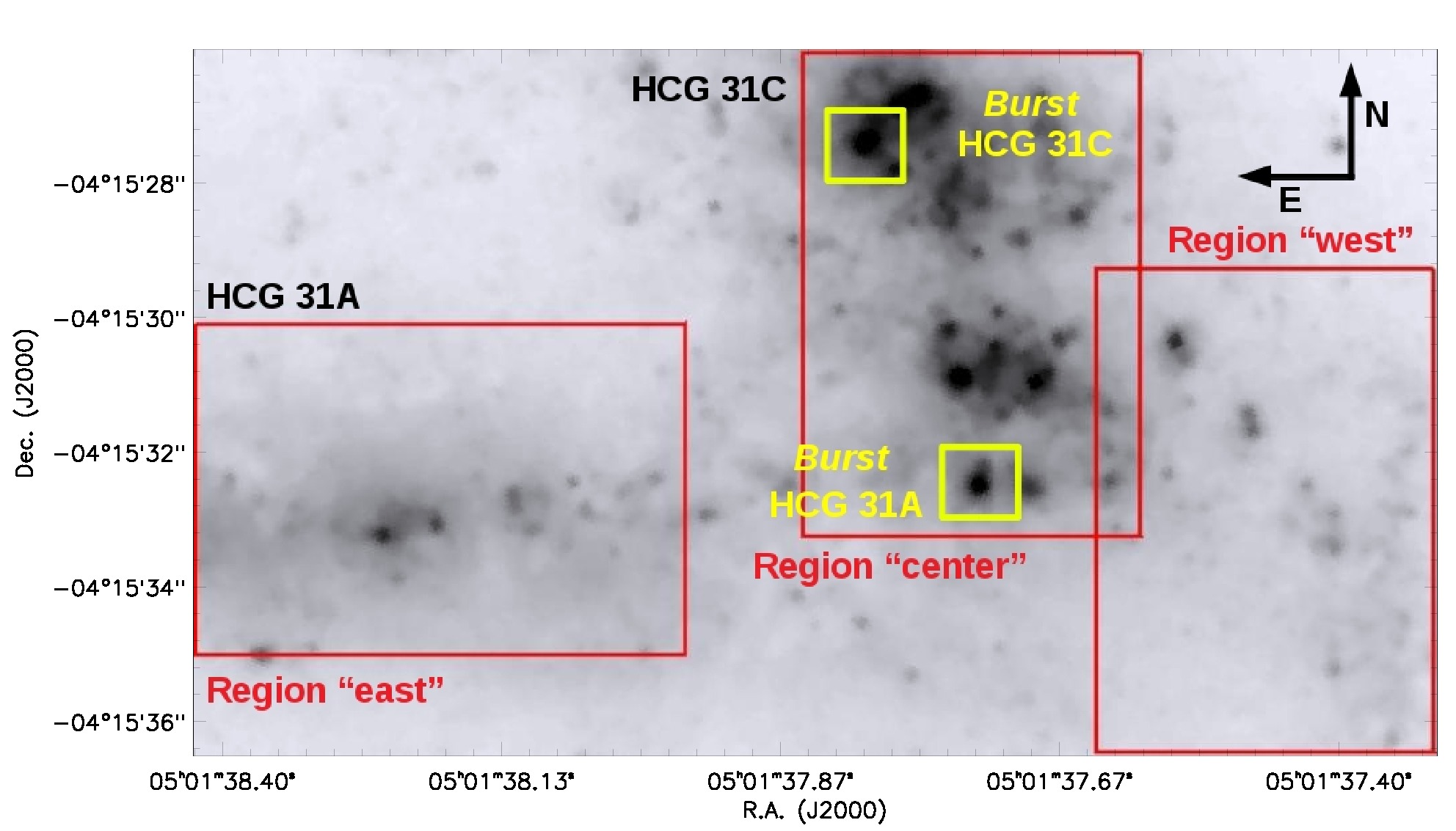}
\caption{\textit{Hubble Space Telescope} (HST) optical image (ACS, F435W) of the system HCG 31A+C.
From left to right: region east, center and west, with a FOV of 5''$\times$7'' each. Red boxes indicate the position of 
the observed IFUs also shown in Figure \ref{hcg31}. Yellow boxes show the positions of the two main bursts. The northern burst
associated to the galaxy HCG 31C and the southern burst to galaxy HCG 31A. Each box has an area of 1.1''$\times$1.1'' 
($\sim$298$\times$298 pc).}
\label{map_hst}
\end{figure*}

\begin{figure*}
\centering \includegraphics[width=2.\columnwidth]{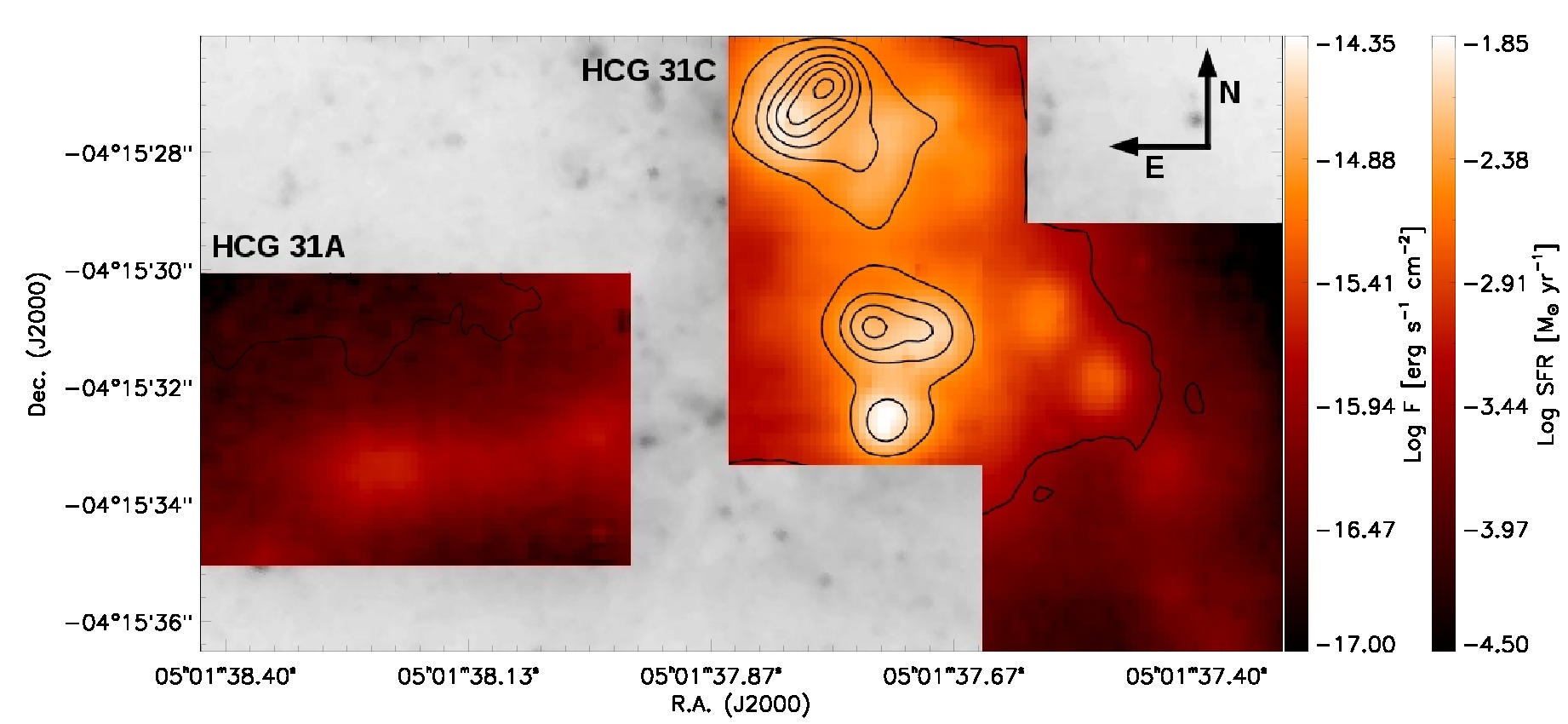}
\centering \includegraphics[width=2.\columnwidth]{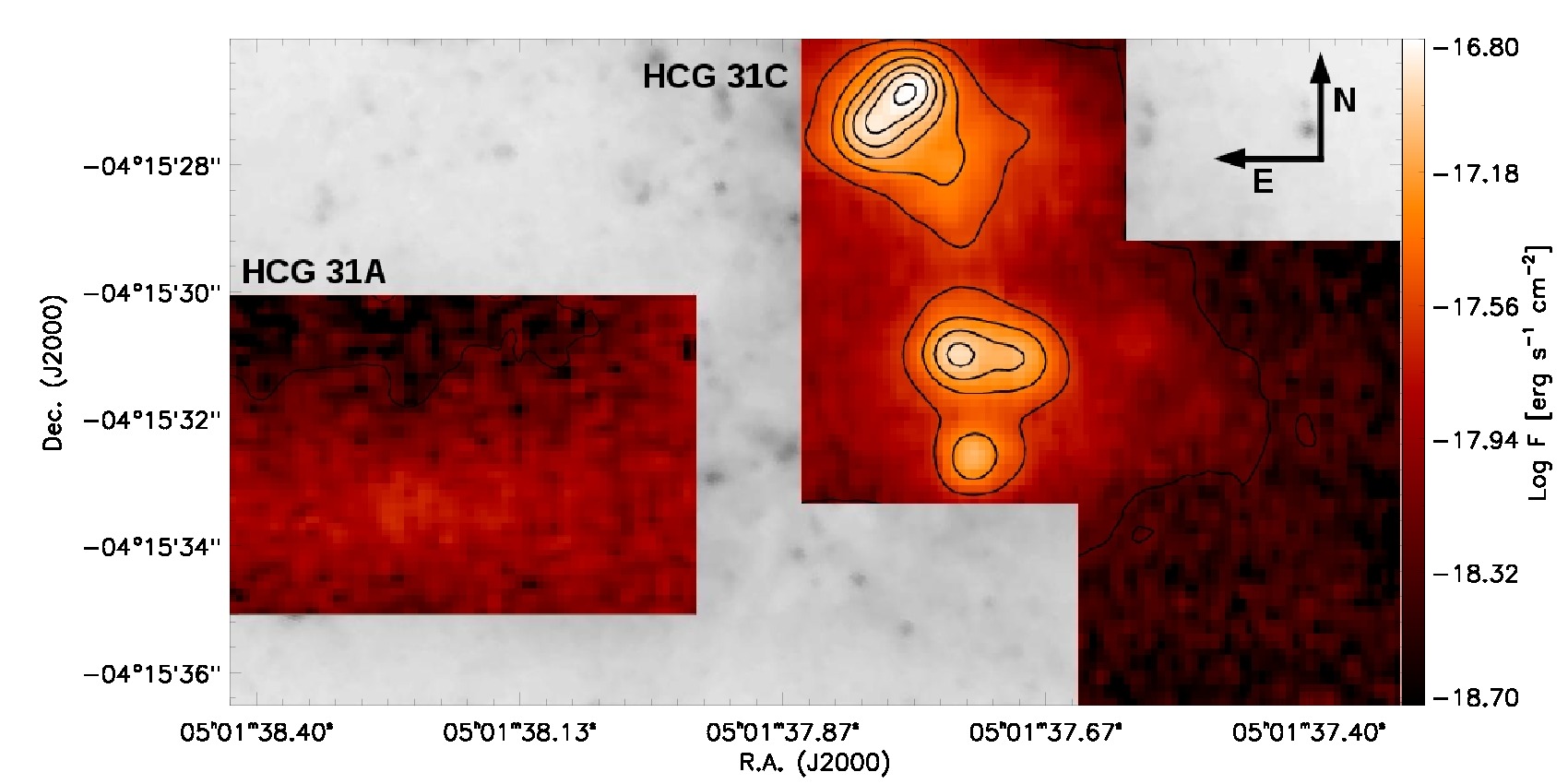}
\caption{Top panel: H$\alpha$ map for the central region of HCG 31A+C. In the same figure we have included a color bar
indicating the estimated SFRs. Bottom panel: Stellar continuum emission for the central region of HCG 31A+C. Both 
emission maps are in units of \mbox{erg s$^{-1}$ cm$^{-2}$} and the black contours represent the stellar continuum emission.}
\label{map_flux}
\end{figure*}

\section{Results}

\begin{figure*}
\centering \includegraphics[width=2.\columnwidth]{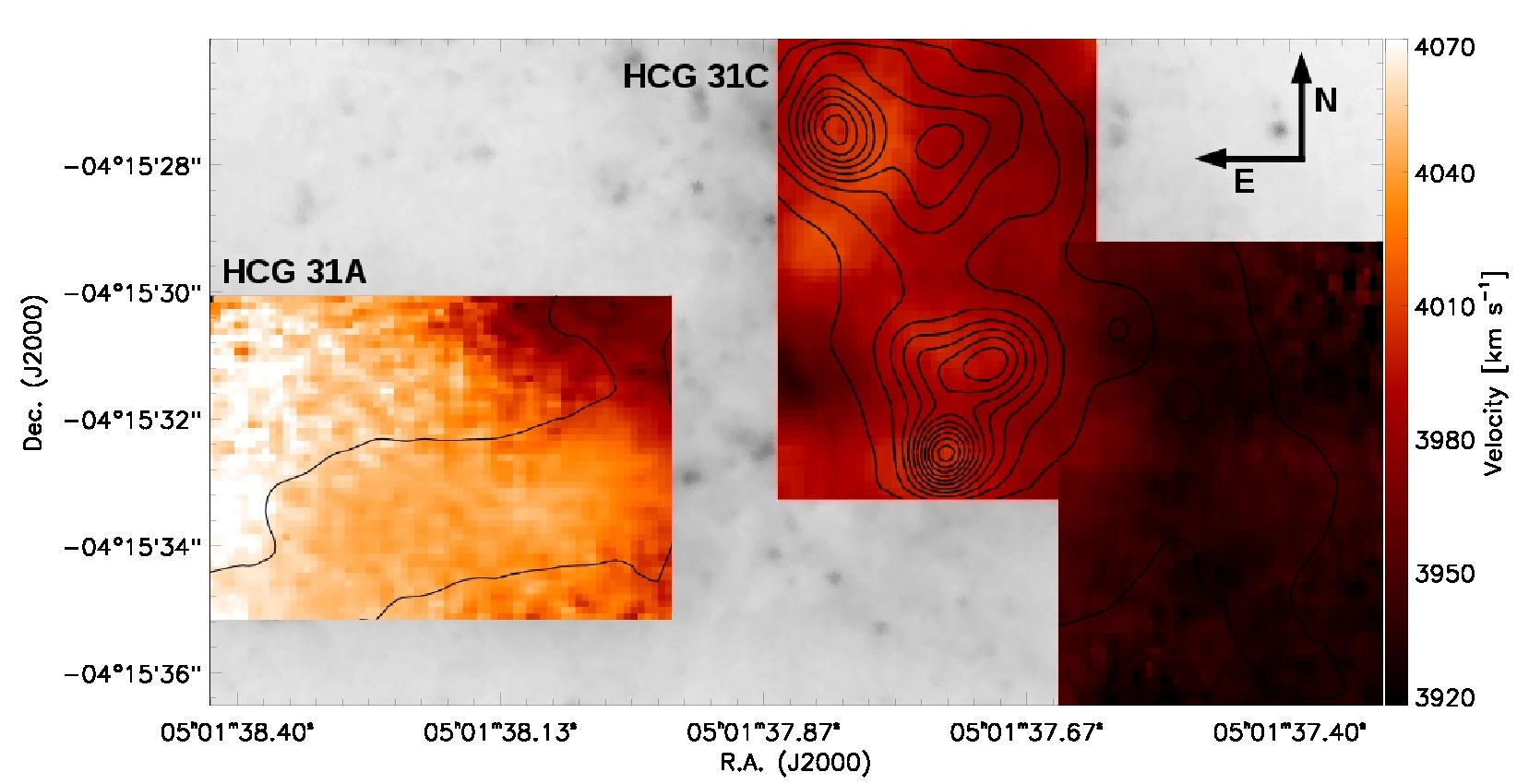}\\
\centering \includegraphics[width=2.\columnwidth]{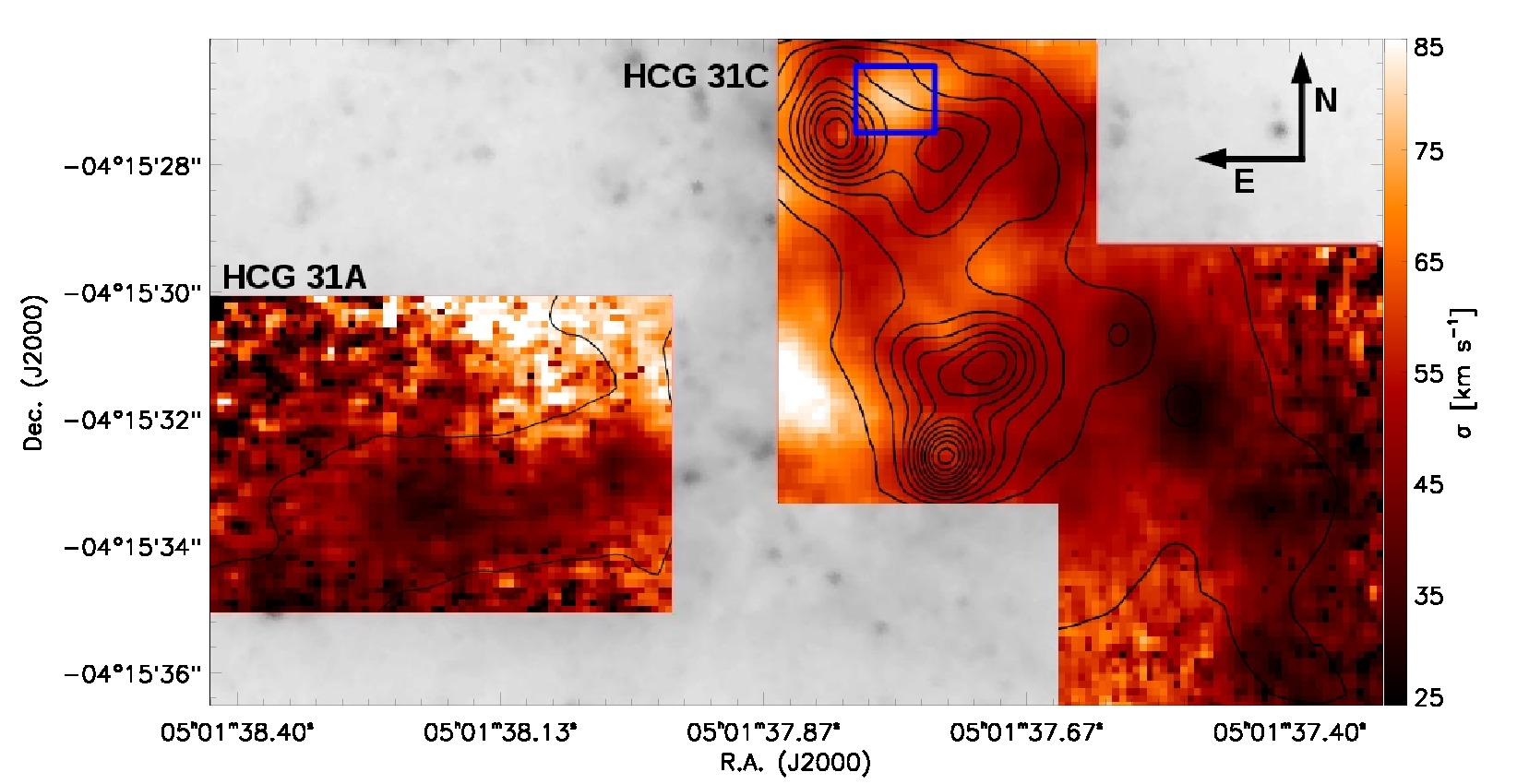}
\caption{Top panel: Velocity map of the central region of HCG 31A+C. Bottom panel: Velocity dispersion map of the central 
region of HCG 31A+C, where the blue box overplotted represent the feature with high velocity dispersion values 
($\sim$70-85\,km\,s$^{-1}$). Black contours overplotted on the maps represent the H$\alpha$ emission.}
\label{map_vel}
\end{figure*}

In Figures \ref{map_flux}, \ref{map_vel}, \ref{map_abundance} and \ref{map_age} we show different maps that we 
derived from the IFU observations of HCG 31A+C. 

\subsection{H$\alpha$ emission and star formation rates.}

In the top panel of Figure \ref{map_flux} we show the H$\alpha$ emission flux map of the central region of HCG 31A+C. In the 
figure, the black contours represent the stellar continuum emission, which was derived from the IFU data and trace the stellar 
emission in this region (see bottom panel of Figure \ref{map_flux}). Inspecting this figure we detect four strong H$\alpha$ 
sources, where the northern objects are linked with the galaxy HCG 31C, while the southern sources belong to the member 
HCG 31A. We also detect diffuse H$\alpha$ emission across the three observed fields. 

The bottom panel of Figure \ref{map_flux} shows the stellar continuum map.
In the case of HCG 31C, there is an offset between the peak of the continuum and the H$\alpha$ emission. The
continuum peaks 1 arcsec ($\sim$270 pc) to the north-west of the maximum in the H$\alpha$ emission. 
This feature has not been previously identified in the central region of HCG 31A+C. This can bring some new insights regarding 
the star formation processes in this merging system. A possible scenario for
this phenomenon is discussed in the next sections. In general our H$\alpha$ map is in agreement with the H$\alpha$ image 
published by V\'ilchez \& Iglesias-P\'aramo (1998). Using our data set we have derived the H$\alpha$ luminosities of the two main bursts 
associated with HCG 31C and HCG 31A, which display luminosities of \mbox{L$_{H\alpha}$=1.89$\times$10$^{41}$ erg s$^{-1}$}
and \mbox{L$_{H_\alpha}$=1.73$\times$10$^{41}$ erg s$^{-1}$}, respectively.
Clearly these values exceed the luminosity of the well known Giant HII regions (like 30 Doradus, which has a luminosity of
\mbox{L$_{H_\alpha}$$\sim$3$\times$10$^{39}$ erg s$^{-1}$}, L\'opez et al. 2014).
The burst detected in HCG 31A and HCG 31C have been studied in the past by Johnson \& Conti (2000), who found that the galaxy 
component AC shows signs of star formation over the past $\sim$10 Myr, and a peak in the EW(H$\alpha$) distribution 
corresponding to star formation about $\sim$5 Myrs ago. In the case of the star-forming complexes associated with HCG 31C,
L\'opez-S\'anchez et al. (2004) detected Wolf-Rayet features, which is consistent with a young stellar population. These
authors also suggest an age of 5 Myrs for the complex associated with HCG 31C, and an age of 7 Myrs for member HCG 31A. 
However, the current analysis allows us to study in detail the physical properties of these bursts, given that we have 
spectroscopic information for each spaxel.

Following Kennicutt (1998), the above H$\alpha$ luminosities were used to estimate the SFRs in the central region of HCG 31A+C.
The right color bar in the top panel of Figure \ref{map_flux} indicates the SFRs in the HCG 31A+C complex. The figure
shows a peak in the SFR located in the HCG 31A complex, which seems to be more compact than the region located
to the north. We estimate a SFR=1.49 M$_{\odot}$yr$^{-1}$ and SFR=1.37 M$_{\odot}$yr$^{-1}$ for the northern and southern 
star-forming regions of HCG 31C and HCG 31A, respectively. Then, for the whole complex, we estimate a 
SFR$\sim$2.86 M$_{\odot}$yr$^{-1}$, which is consistent with the value derived by L\'opez-S\'anchez et al. (2004) for the
region HCG 31A+C (see their Table 9).
When we derived the SFR for each observed field (``east'', ``center'' and ``west''), we found that fields ``east'' and ``west''
display a modest SFR, when compared with the SFR derived for the central field (0.69, 9.16 and 1.21 M$_{\odot}$yr$^{-1}$, 
respectively). All these values are listed in Table 1 including the uncertainties associated to these measurements.
Integrating the SFR for all the observed members of HCG 31A+C, we 
determine a SFR=11.06 M$_{\odot}$yr$^{-1}$. Gallagher et al. (2010) used UV and IR information to derive the SFR of all members
of HCG 31, finding a value of $\sim$10.6 M$_{\odot}$yr$^{-1}$. They note that the main contribution in the SFR comes from
members A+C and E. Some possible reasons for the discrepancies between both determinations are mentioned in the \S 
\ref{properties}.

\begin{table*}
\centering
\scriptsize
\caption{Luminosities and SFRs of HCG 31A+C}
\begin{threeparttable}
        {\small
       \begin{tabular}{lcccccc}
\hline
Source              & L$_{H\alpha}$$^{(1)}$$^*$       & L$_{H\alpha}$$^{(2)}$$^*$       & L$_{H\alpha}$$^{(3)}$$^*$       & SFR$^{(1)}$$^{**}$   & SFR$^{(2)}$$^{**}$   & SFR$^{(3)}$$^{**}$\\
                    & $\times$10$^{41} $erg s$^{-1}$  & $\times$10$^{41} $erg s$^{-1}$  & $\times$10$^{41} $erg s$^{-1}$  & M$_{\odot}$yr$^{-1}$ & M$_{\odot}$yr$^{-1}$ & M$_{\odot}$yr$^{-1}$ \\ 
  \hline
Burst HCG 31C$^a$   &  1.43 $\pm$ 0.13                &  1.55 $\pm$ 0.14                &  1.89 $\pm$ 0.18                &  1.13 $\pm$ 0.11     &  1.22 $\pm$ 0.11     &  1.49 $\pm$ 0.14 \\
Burst HCG 31A$^a$   &  1.31 $\pm$ 0.12                &  1.42 $\pm$ 0.13                &  1.73 $\pm$ 0.16                &  1.03 $\pm$ 0.10     &  1.12 $\pm$ 0.10     &  1.37 $\pm$ 0.13 \\
HCG 31 East         &  0.63 $\pm$ 0.06                &  0.70 $\pm$ 0.65                &  0.87 $\pm$ 0.08                &  0.50 $\pm$ 0.05     &  0.55 $\pm$ 0.05     &  0.69 $\pm$ 0.06 \\
HCG 31 Center       &  8.77 $\pm$ 0.82                &  9.50 $\pm$ 0.89                & 11.60 $\pm$ 1.08                &  6.93 $\pm$ 0.65     &  7.51 $\pm$ 0.70     &  9.16 $\pm$ 0.85 \\
HCG 31 West         &  1.16 $\pm$ 0.11                &  1.26 $\pm$ 0.17                &  1.53 $\pm$ 0.14                &  0.92 $\pm$ 0.09     &  0.99 $\pm$ 0.09     &  1.21 $\pm$ 0.11 \\
HCG 31 All          & 10.56 $\pm$ 0.83                & 11.46 $\pm$ 1.12                & 14.00 $\pm$ 1.10                &  8.35 $\pm$ 0.66     &  9.05 $\pm$ 0.71     & 11.06 $\pm$ 0.86 \\
\hline
\label{tabla1}
\end{tabular} 
}
\begin{tablenotes}
\item[\emph{$^a$}]{Considering a box of 1.1''$\times$1.1'', corresponding to 298$\times$298 pc.}
\item[\emph{$^*$}]{Estimated using the integrated H${\alpha}$ emission flux.}
\item[\emph{$^{**}$}]{Using the measured integrated L$_{H\alpha}$ luminosity as input to Kennicutt's (1998) formula (2), and assuming continuous star formation.  }
\item[\emph{$^{(1)}$}]{Value before extinction correction.}
\item[\emph{$^{(2)}$}]{Value corrected by Galactic extinction using the Fitzpatrick extinction law.}
\item[\emph{$^{(3)}$}]{Value corrected by Galactic extinction using the Fitzpatrick extinction law, and internal extinction using the Calzetti extinction law with a nebular color excess of E(B-V)$_{gas}$=0.08 mag.}
\end{tablenotes} 
\end{threeparttable}
\end{table*}

\subsection{Kinematics: Velocity fields and velocity dispersion maps}
\label{kinematics}

In the top panel of Figure \ref{map_vel} we show the velocity field of the central region of HCG 31. The uncertainty 
associated to the radial velocities is $\sim$10\,km\,s$^{-1}$, which was estimated adding in quadrature the error in the 
wavelength calibration and the FWHM. We found that the main
star-forming bursts in HCG 31A and HCG 31C display similar radial velocities, which reach values of $\sim$4000\,km\,s$^{-1}$.
Although we have a small field of view in our observations, we observe a perturbed velocity field. We can not identify a 
kinematic major axis. This fact suggests that non-circular motions are present in the central region of HCG 31, which is 
expected, given the strong signatures of interactions. In general, the velocity field presented in Figure \ref{map_vel} is in 
agreement with the kinematic analysis presented by Amram et al. (2007). 

The velocity dispersion map for HCG 31 is presented in the bottom panel of Figure \ref{map_vel}. The uncertainty associated
to the velocity dispersion is of 13\,km\,s$^{-1}$. The two main bursts of star formation, belonging to HCG 31A and HCG 31C, 
display velocity dispersions of $\sim$55\,km\,s$^{-1}$. In this figure, the most interesting feature corresponds to a region 
located in HCG 31C, which displays quite high velocity dispersions $\sim$85\,km\,s$^{-1}$, which is identified by a blue
box. The velocity dispersion in this region differs by at least 1$\sigma$ (13\,km\,s$^{-1}$) from that of its 
neighborhood, as calculated by considering the largest variation across the field, in an area of 5''$\times$ 7''
($\sim$10\,km\,s$^{-1}$). The variation across an area of 2''$\times$ 5'', in the central region where the feature is located,
is smaller ($\sim$6\,km\,s$^{-1}$).
Considering this, we can suggest the existence of a peak in the velocity dispersion located in this area.
This source lies at the same spatial position as the peak in the continuum contours (see bottom panel of Figure 
\ref{map_flux}). Given the medium spectral resolution of our observations, this broadening can emerge as the sum of, at least,
two components of the H$\alpha$ emission line. These components may be associated with ionized gas that is being blown out by 
a star cluster, which can be identified with the continuum information. We can now speculate that the offset between the
H$\alpha$ and continuum emission at this 
location (star-forming burst in HCG 31C) is produced by a stellar cluster, which is cleaning its environment. This process may
be triggering star formation in its neighborhood, and that is the reason why we observe a strong H$\alpha$ emitting source
close to the continuum peak. This kind of phenomenon has been studied in local giant star-forming regions, such as N11
(Walborn \& Parker 1992, Barb\'a et al. 2003) and in the H{\sc ii} region NGC\,604 in M33 (Ma\'iz-Apellaniz et al. 2004, 
Tosaki et al. 2007, Barb\'a et al. 2009, Fari\~na et al. 2012, Mart\'inez-Galarza et al. 2012).

\begin{figure*}
\includegraphics[width=2.\columnwidth]{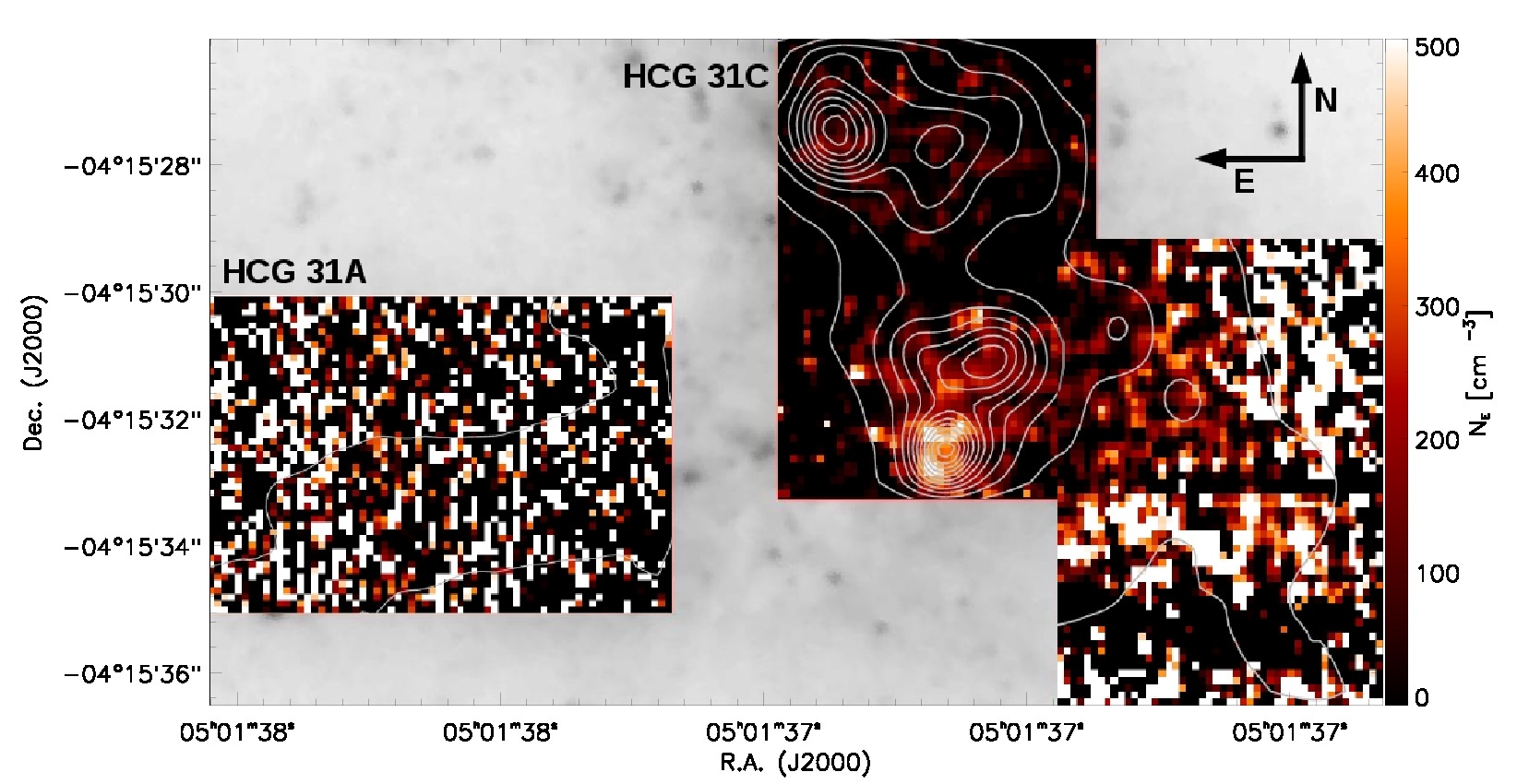}\\
\includegraphics[width=2.\columnwidth]{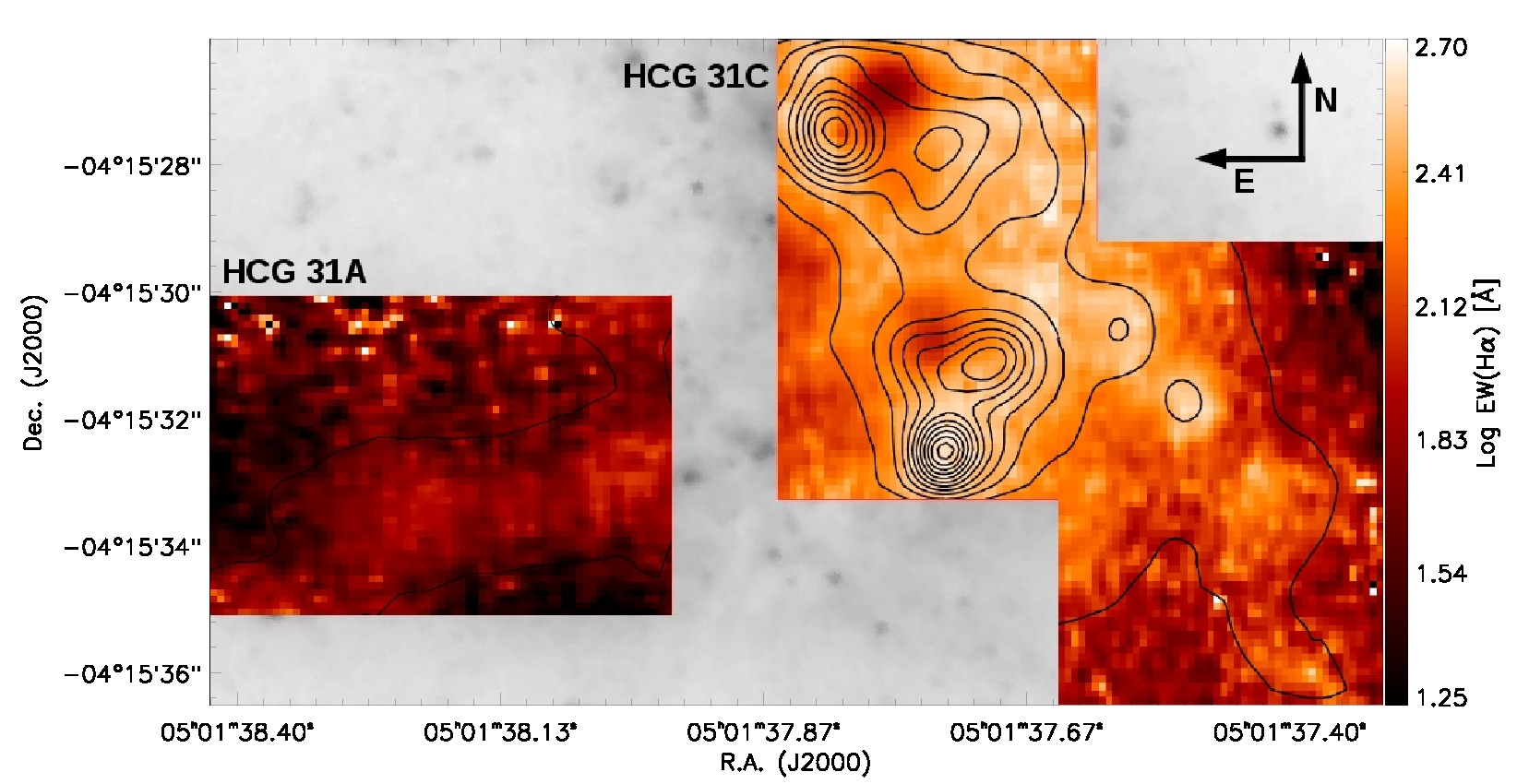}
\caption{Top panel: Electron density map for the central region of HCG 31A+C. Bottom panel: H$\alpha$ equivalent width map for
the central region of HCG 31A+C. Black and white contours overplotted on the maps represent the H$\alpha$ emission.}
\label{map_abundance}
\end{figure*}

\subsection{Oxygen abundances and electron Densities}\label{res_abundance}

The oxygen abundance map of the central region of HCG 31A+C has been recently published by our group (Torres-Flores et al. 2015).
This map was derived by using the N2 index, which is defined as:
\begin{equation}\label{N2}
 \textrm{N2}=log([\textrm{N}\textsc{ii}]\lambda6584/\textrm{H}\alpha),
\end{equation}
and using the Marino et al. (2013) calibration. Torres-Flores et al. (2015) found a smooth metallicity 
gradient between HCG 31A and HCG 31C (\mbox{1.5$\times$10$^{-4}$\,dex\,pc$^{-1}$}).
Inspecting high-resolution Fabry-Perot data published by
Amram et al. (2007), they found a gas flow between both galaxies, which confirms the scenario of metal mixing produced by gas
flows, given the smooth gradient between HCG 31A and HCG 31C.

In the top panel of Figure \ref{map_abundance} we show the electron density map of HCG 31A+C, where the contours 
represent the H$\alpha$ emission. Given the low intensity of the [SII]$\lambda$$\lambda$6717,6731\,{\AA} emission lines, most
of the information comes from the main star-forming burst.
The burst associated to HCG 31A displays an average electron density of \mbox{N$_{e}$=228$\pm$120 cm$^{-3}$}, with a peak
of \mbox{N$_{e}\sim$500 cm$^{-3}$}. In the case of the galaxy HCG 31C, it displays an average of 
\mbox{N$_{e}$=82$\pm$50 cm$^{-3}$}, reaching a peak of \mbox{N$_{e}\sim$150 cm$^{-3}$}. L\'opez-S\'anchez et al. (2004)
estimated the electron density for the different components of HCG 31. In the case of HCG  31C, these authors found a value of
\mbox{N$_{e}$=270$\pm$70 cm$^{-3}$} (the value for the HCG 31A was not estimated). However, we note that our data allow us to 
obtain the electron density for each spaxel in HCG 31A+C. Therefore, this fact can affect any comparison with previous results,
which mainly have been obtained from longslit spectroscopy.

\subsection{H$\alpha$ Equivalent Width and age estimation}
\label{wha}

Given that the H$\alpha$ emission traces recent star formation, the EW(H$\alpha$) can be used to estimate the ages of these 
events (e.g., Leitherer \& Heckman 1995). The H$\alpha$ Equivalent Width is important since it gives us an estimation
of the ratio between the ionizing photons coming from massive stars and the continuum photons from the stellar population and gas
(Cedr\'es, Cepa \& Tomita 2005).
The bottom panel of Figure \ref{map_abundance} shows EW(H$\alpha$) values for the central region of HCG 31A+C, where the 
contours represent the H$\alpha$ emission. The EW(H$\alpha$) varies between 18\,{\AA}$<$EW(H$\alpha$)$<$501\,{\AA}.
For the two main bursts we have estimated a mean value of \mbox{EW(H$\alpha$)= 338$\pm$50\,{\AA}} for burst in HCG 31A and of
\mbox{EW(H$\alpha$)= 238$\pm$90\,{\AA}} for burst in HCG 31C, whose values are similar to those measured in the star forming
complexes in the ``west'' region.


\begin{figure*}
\centering \includegraphics[width=2.\columnwidth]{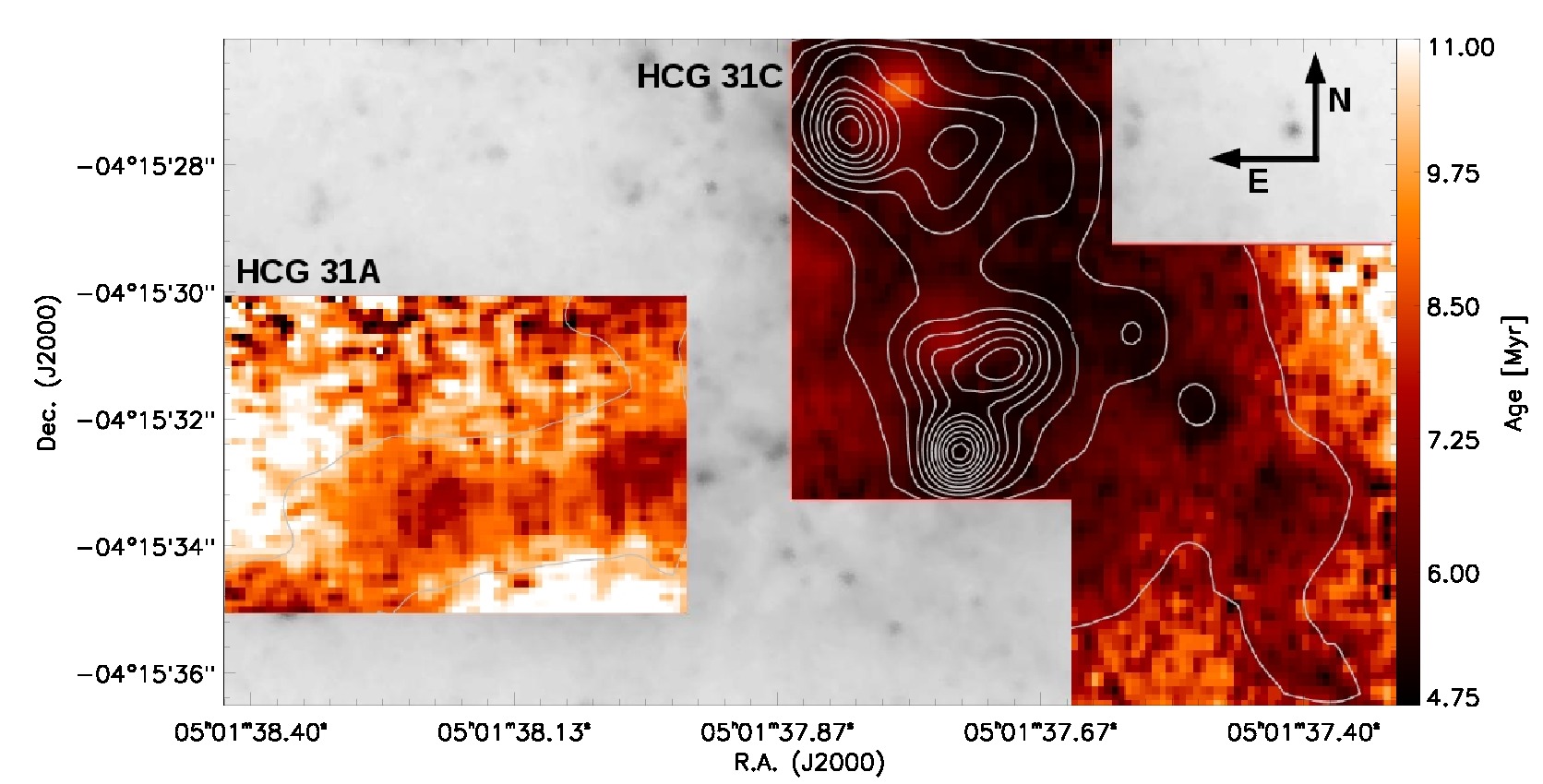}\\
\centering \includegraphics[width=2.\columnwidth]{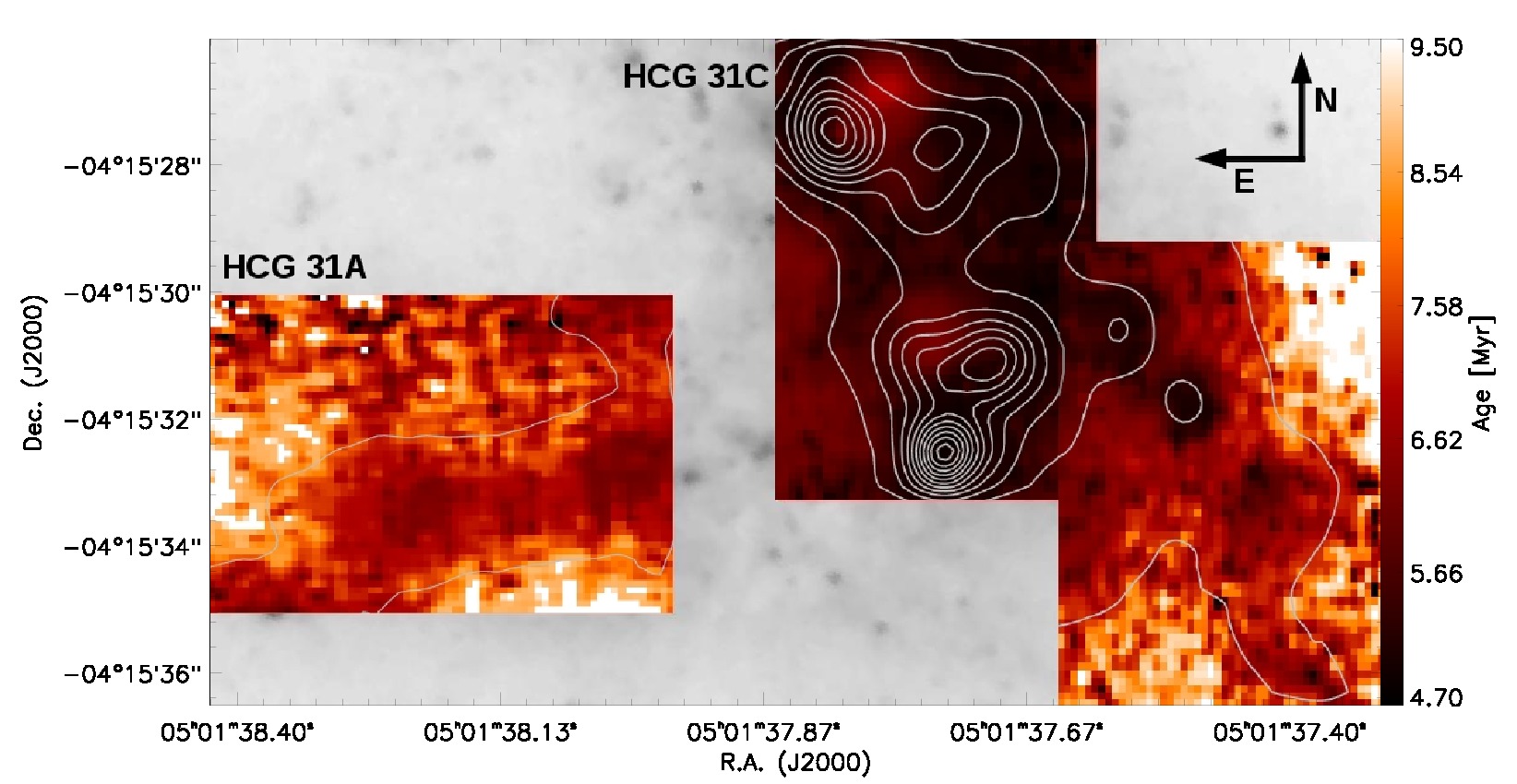}
\caption{Top panel: Age map estimated assuming an instantaneous star formation law for the central region of HCG 31A+C for a
metallicity of Z=0.004. Bottom panel: similar to Top Panel, but for Z=0.008. Both cases for a Salpeter initial mass function
(IMF) with a mass limit of 1-100 M$_\odot$. In both panels white contours overplotted on the map represent the H$\alpha$ 
emission.}
\label{map_age}
\end{figure*}

From the EW(H$\alpha$) values we have estimated the age of the star-forming objects located in the central region of the
system HCG 31A+C. These maps are shown in Figure \ref{map_age}. In the top an bottom panels the ages were estimated by assuming
a metallicity of Z=0.004 and Z= 0.008, respectively. In both maps the contours represent the H$\alpha$ emission.
Inspecting the top panel (Z=0.004), we can observe ages ranging from 5 to 11 ($\pm$2) Myrs, through the ``east'', 
``center'' and ``west'' regions. In the bursts associated to the galaxies HCG 31C and HCG 31A, we have estimated a mean age of 6$\pm$2 
and 5$\pm$1 Myrs, respectively.
Considering the uncertainties, we can suggest that both bursts have similar ages. The bottom panel in Figure \ref{map_age}
(Z=0.008) shows similar ages with small variations, but following the same trend with ages between 5--10 ($\pm$2) Myrs.  

We have also estimated the age at the same location of the stellar continuum emission peak (see bottom panel Figure \ref{map_flux})
where we obtain an age of 9 Myrs, while the H$\alpha$ emission peaks in the burst associated to HCG 31C display an age of 5 
Myrs (see top panel Figure \ref{map_flux}). This result is in agreement with the scenario we proposed in 
\S \ref{kinematics}, where the stellar continuum emission is coming from a stellar cluster ($\sim$9 Myrs) which would be
cleaning the ISM triggering star formation around it ($\sim$5 Myrs).
The lowest ages are found in the galaxy HCG 31C and (even considering the uncertainties) are consistent with
Wolf-Rayet stars. This fact is consistent with the spectrum of this region published by L\'opez-S\'anchez et al. (2004). 

Although previous works have already dated some star-forming regions in HCG 31, the IFU data allow  estimating the ages of 
different structures in a spatially-resolved way, like in the star-forming regions in HCG 31A and HCG 31C, and the ``east'', 
``center'' and ``west'' regions. Also, the current results show that the region where the super stellar cluster candidate lies is older
than the burst associated with the galaxy HCG 31C. This allows us to add a
new piece of evidence to the scenario we proposed where we have a stellar cluster, which is cleaning its environment triggering
star formation. On the other hand, other authors have estimated the ages for a more global area in this system
(e.g.  L\'opez-S\'anchez et al. 2004, Johnson \& Conti 2000) which range from 5 to 10\,Myrs.

\begin{figure}
   \centering
        \includegraphics[width=\columnwidth]{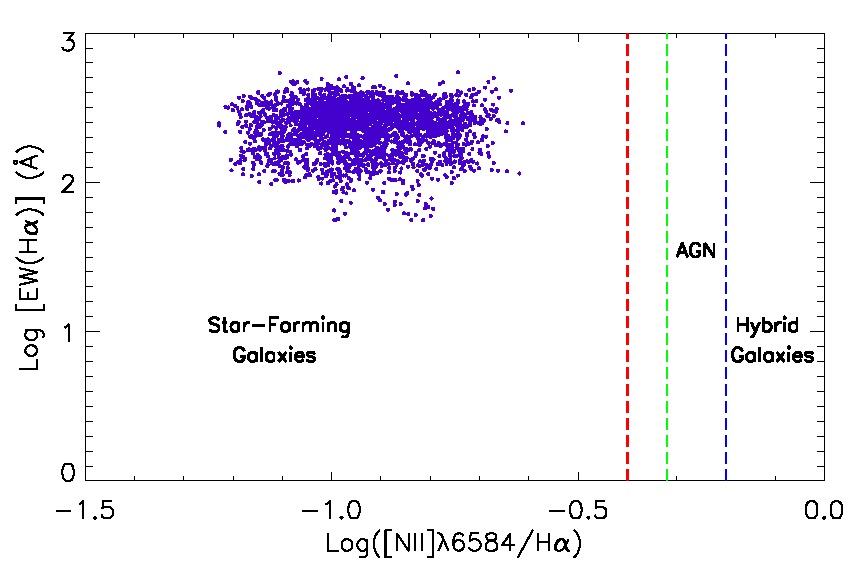}
        \caption{\small{Diagnostic diagram EW(H$\alpha$) versus N2. The values presented belong to the region ``center''.
        The limits proposed by Stasi\'nska et al. (2006) for the galaxy classification are: red dashed line for star-forming
        galaxies, with \mbox{-0.4 $<$log([N{\sc ii}]/H$\alpha$)}; the region between the red and blue dashed lines correspond
        to AGNs, with \mbox{-0.4$<$log([N{\sc ii}]/H$\alpha$)$<$-0.2}; and blue dashed line shows the limit for hybrid galaxies
        with \mbox{log([N{\sc ii}]/H$\alpha$)$>$-0.2}. Green dashed line represent the limit proposed by Kauffmann et al. (2003)
        to distinguish between star-forming galaxies and AGNs, corresponding to \mbox{-0.32$<$log([N{\sc ii}]/H$\alpha$)}.}}
        \label{ionizing_mechanism}
\end{figure}

\subsection{Ionizing mechanism}

In order to analyze the ionizing mechanism in the ``center'' region of HCG 31A+C (where most of the H$\alpha$ emission 
arises) we have included an EW(H$\alpha$) versus N2 (log([N{\sc ii}]/H$\alpha$)) diagnostic diagram, which is presented in 
Figure \ref{ionizing_mechanism}. 
This kind of diagram was proposed for the first time by Cid Fernandes et al. (2010), who used it for a gas ionization mechanism
classification of their galaxy sample, considering the limits proposed by Kewley et al. (2001), Kauffmann et al. (2003) and 
Stasi\'nska et al. (2006).  Cid Fernandes et al. (2010) point out the utility of these diagrams, which arises from the
information we can get from both physical parameters (EW and N2). First, they mention that the line ratio can give 
information of the ionized gas physical conditions, and the EW(H$\alpha$) provides an estimation of the ionization power of 
the stellar population. Kauffmann et al. (2003) present a classification based on the [N{\sc ii}]/H$\alpha$ ratio, proposing an
upper limit of \mbox{-0.32$<$log([N{\sc ii}]/H$\alpha$)} to star-forming galaxies (green dashed line in Figure 
\ref{ionizing_mechanism}).
Stasi\'nska et al. (2006) present techniques to distinguish star-forming galaxies from active galactic nuclei galaxies (AGN). 
One of these is also based on the [N{\sc ii}]/H$\alpha$ ratio, where the classification is given by:
\mbox{-0.4$<$log([N{\sc ii}]/H$\alpha$)} for star-forming galaxies (red dashed line in Figure \ref{ionizing_mechanism}); 
\mbox{-0.4$<$log([N{\sc ii}]/H$\alpha$)$<$-0.2} for AGNs (region between red and blue dashed line in Figure
\ref{ionizing_mechanism}); and \mbox{log([N{\sc ii}]/H$\alpha$)$>$-0.2} for hybrid galaxies (blue dashed line in Figure 
\ref{ionizing_mechanism}). In these diagrams, the limits proposed for the galaxy classification given by Kauffmann et al. (2003)
and Stasi\'nska et al. (2006) are approximately vertical. This fact allows us to carry out this classification considering 
the [N{\sc ii}]/H$\alpha$ ratio (Cid Fernandes et al. 2010).

In Figure \ref{ionizing_mechanism} we note that the N2 values for HCG 31A+C are in the range of 
\mbox{-1.2 $<$ log([N{\sc ii}]/H$\alpha$) $<$ -0.5} (purple points), while the EW(H$\alpha$) values are in
\mbox{1.3 $<$ log(EW(H$\alpha$)) $<$ 2.7}.
In this diagram we can observe that the ``center'' region is placed over a region where the gas in the galaxies is ionized by star
formation. This corresponds to the limits proposed by Kauffmann et al. (2003) (green dashed line) and Stasi\'nska et al. (2006) 
(red dashed line). From these results we conclude that the ionizing mechanism in HCG 31A+C is star formation, excluding 
AGN activity.

\subsection{Intensity versus $\sigma$ plots}
\label{sigmavsintensity}

\begin{figure*}
\centering \includegraphics[width=2.\columnwidth]{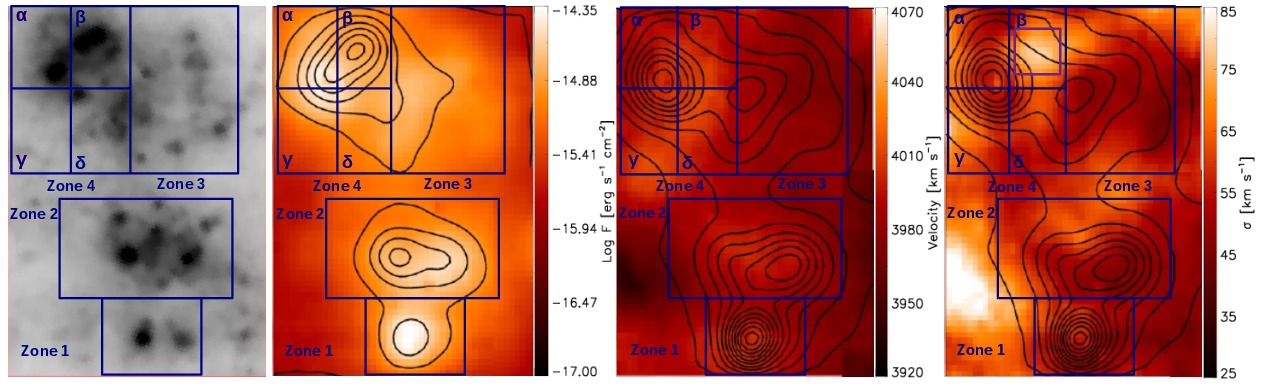}
\caption{\small{From left to right, region ``center'' from the optical {\sc hst} image (from Figure \ref{map_hst}), H$\alpha$ 
flux map (contours: stellar continuum emission) (from Figure \ref{map_flux}), velocity field map and velocity dispersion
map (both from Figure \ref{map_vel}). The blue squares in panels show the four zones with the highest H$\alpha$ emission in the
region ``center'', denominated as Zone 1, Zone 2, Zone 3 and Zone 4. Zona 4 is subdivided in the sub-zones $\alpha$, $\beta$,
$\gamma$ and $\delta$. The purple square overplotted in the last panel is located in the position of the suggested {\it shell}
in this region, with an area of 0.58''$\times$0.58'' (158$\times$158 pc).}}
\label{mapas_diagrama}
\end{figure*}

In section \ref{kinematics} we suggest the presence of a super stellar cluster in HCG 31C, which would be cleaning its 
enviroment and triggering star formation in its vicinity. This suggestion was supported by the measured velocity dispersion 
at that location. This kind of phenomena has been reported for other systems, but for the first time, we suggest these events
are taking place in the main central region of HCG 31A+C, in the interface area between the galaxies HCG 31A and HCG 31C.
In order to investigate the presence of this (or maybe also other) structure and the plausibility of a possible
wind-triggered star-forming scenario we have divided the region ``center'' in four zones (considering the major H$\alpha$ 
emission regions) and we present ``$\sigma$'' versus ``flux'' diagrams for these zones. These are represented by blue
squares in the four panels of Figure \ref{mapas_diagrama}, which are enumerated respectively as Zone 1, Zone 2, Zone 3 and
Zone 4. In this figure, from left to right, the panels correspond to the region ``center'' extracted from the optical {\sc hst}
image (Fig. \ref{map_hst}), H$\alpha$ emission map (Fig. \ref{map_flux}) with the stellar continuum emission represented by
the black contours, velocity field map (Fig. \ref{map_vel}) and the velocity dispersion map (Fig. \ref{map_vel}), where the 
contours represent the H$\alpha$ emission. In the last panel of Figure \ref{mapas_diagrama} we have overplotted a purple square
which represents the super shell produced by the suggested super stellar cluster in this region, with an area of 
0.58''$\times$0.58'' (158$\times$158 pc). 

\begin{figure*}
\centering \includegraphics[width=\columnwidth]{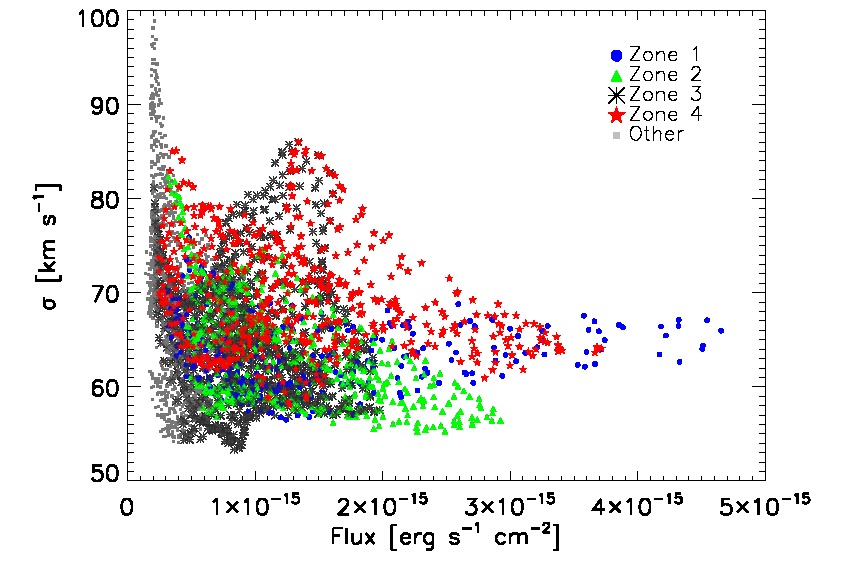}\\
\centering \includegraphics[width=\columnwidth]{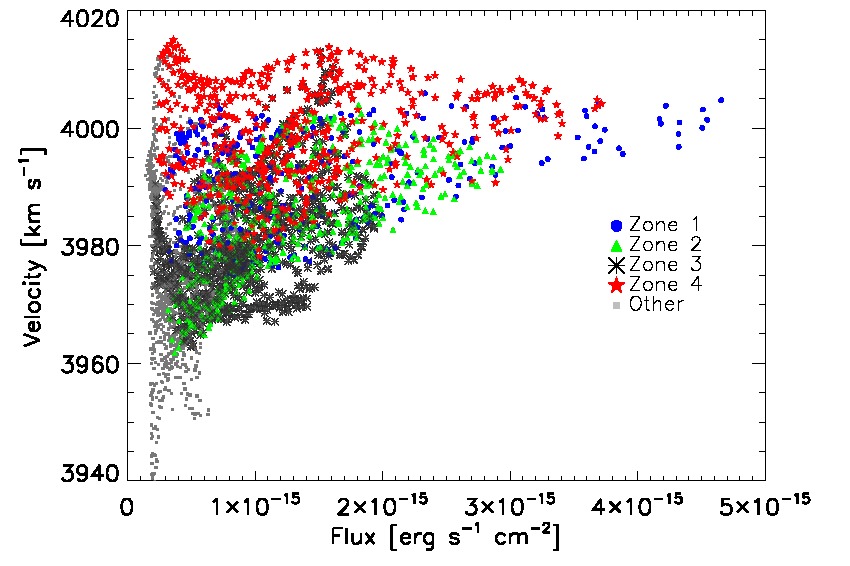}\\
\centering \includegraphics[width=\columnwidth]{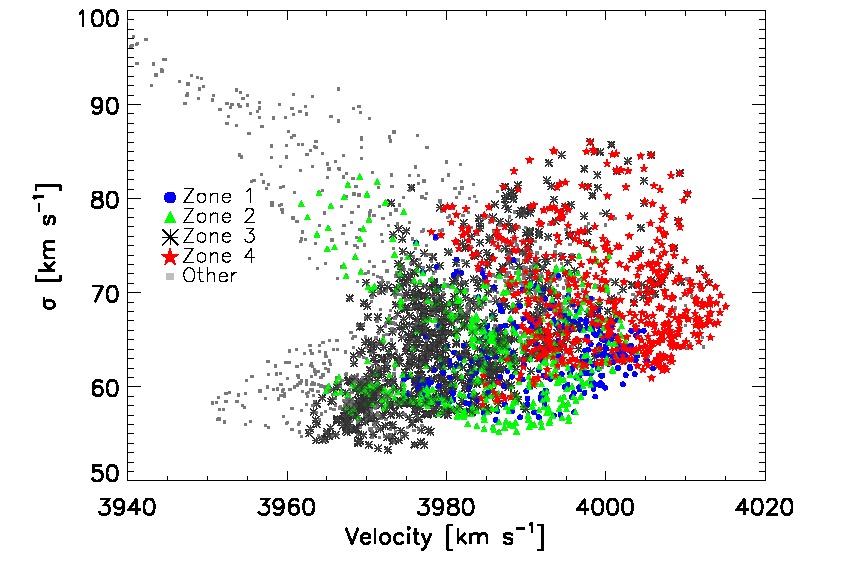}\
\caption{\small{Top panel: $\sigma$ versus flux diagram. Middle panel: radial velocity versus flux diagram. 
Bottom panel: $\sigma$ versus radial velocity diagram. The diagrams consider the zones 1, 2, 3 and 4, which are represented by
blue, green, black and red symbols, respectively. Grey symbols represent the points that do not belong to any of these zones.}}
\label{diagrama}
\end{figure*}

The ``$\sigma$'' versus ``flux'' diagnostic diagrams are a useful tool to identify the emission line profile broadening 
mechanism (i.e: Mu\~noz-Tu\~n\'on et. al 1996, Mart\'inez-Delgado et al. 2007, Moiseev \& Lozinskaya 2012).
In the top panel in Figure \ref{diagrama} we present the ``sigma'' versus ``flux'' diagram for the four zones in region 
``center'', while in middle and bottom panels we include the ``radial velocity'' versus ``flux'' and ``$\sigma$'' versus
``radial velocity'' diagrams, respectively. In the three diagrams, the blue, green, black and red symbols correspond to the 
zones 1, 2, 3 and 4, respectively. Grey symbols represent the points which do not belong to any of these zones.
Analyzing the ``$\sigma$'' versus ``flux'' diagram we can see that the Zone 4 (red symbols) shows a different spatial
distribution in comparison with the others regions, reaching the highest velocity dispersions of the region ``center'' 
($\sigma\sim$60-85 km s$^{-1}$) and fluxes up to \mbox{$\sim$3.7$\times$10$^{-15}$ ergs s$^{-1}$ cm$^{-2}$}. 

This difference is also observed in the ``radial velocity'' versus ``flux'' (middle panel) and ``$\sigma$'' versus ``radial 
velocity'' (bottom panel) diagrams. This zone displays a wide velocity range between 3985-4015 km s$^{-1}$, with predominance of 
high velocity values. This fact could indicate the existence of several structures in the region 
``center'', where each structure shows a different kinematic behavior. This could suggest that the H$\alpha$ 
line profile broadening mechanism in Zone 4 is different to the other zones. 
On the other hand, from the position of the gray symbols in the ``$\sigma$'' versus ``flux'' diagram (top panel Figure \ref{diagrama}),
we can say these do not belong to any knot or do not form any shell. These are part of the envelope that surrounds H{\sc ii}
regions.

\begin{figure*}
\centering \includegraphics[width=\columnwidth]{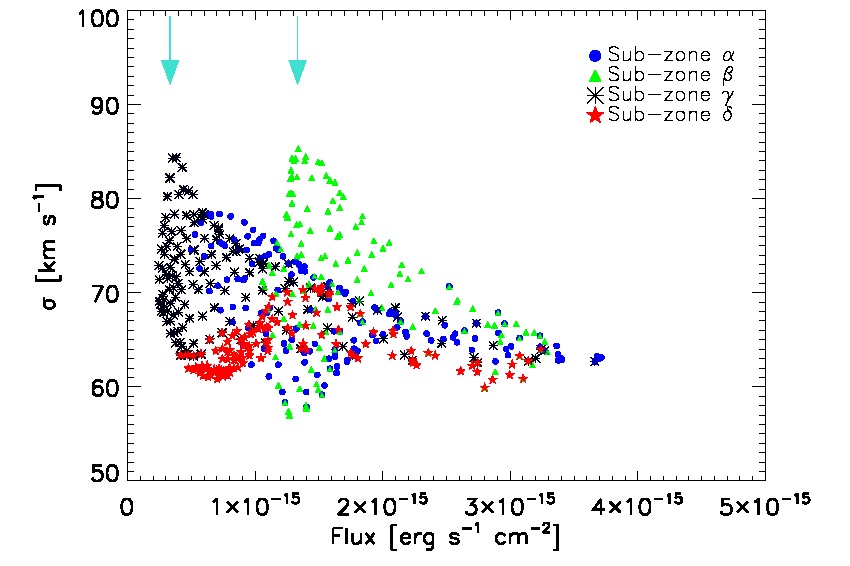}\ \includegraphics[width=\columnwidth]{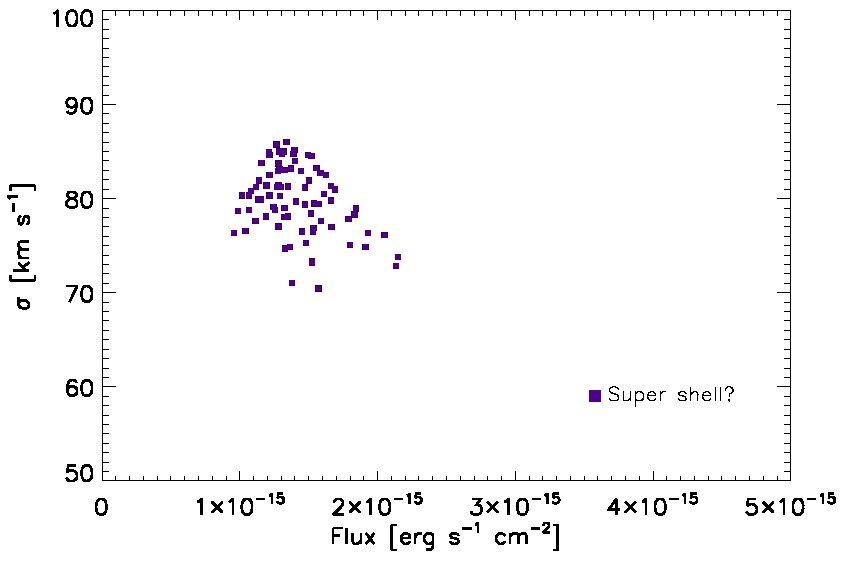}\\
\includegraphics[width=\columnwidth]{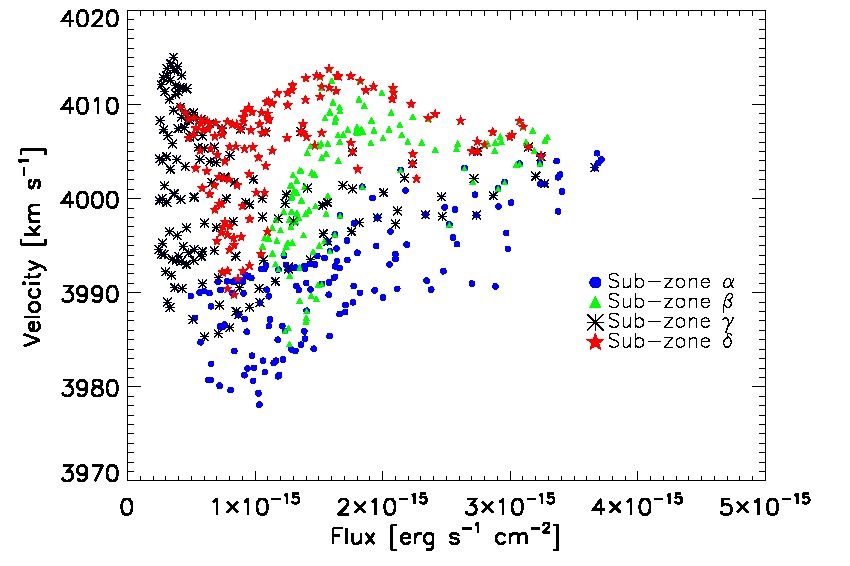}\ \includegraphics[width=\columnwidth]{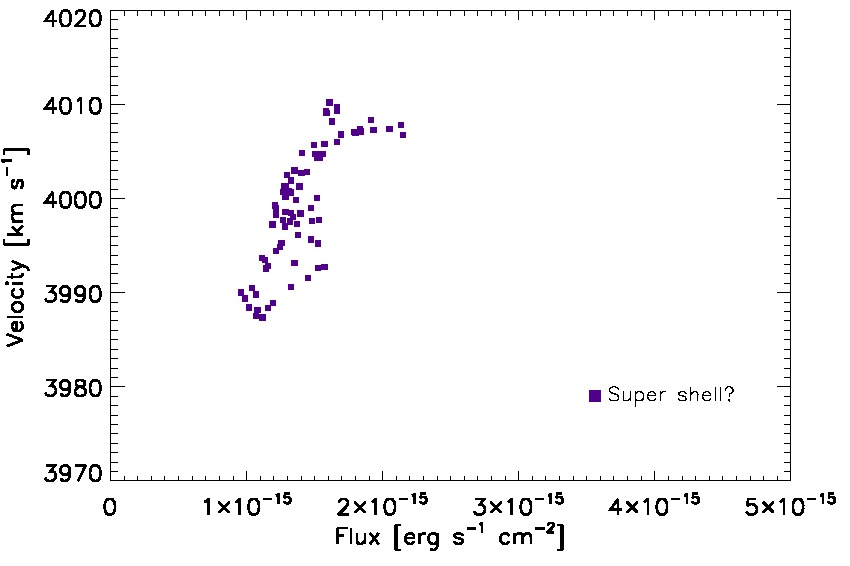}\\
\includegraphics[width=\columnwidth]{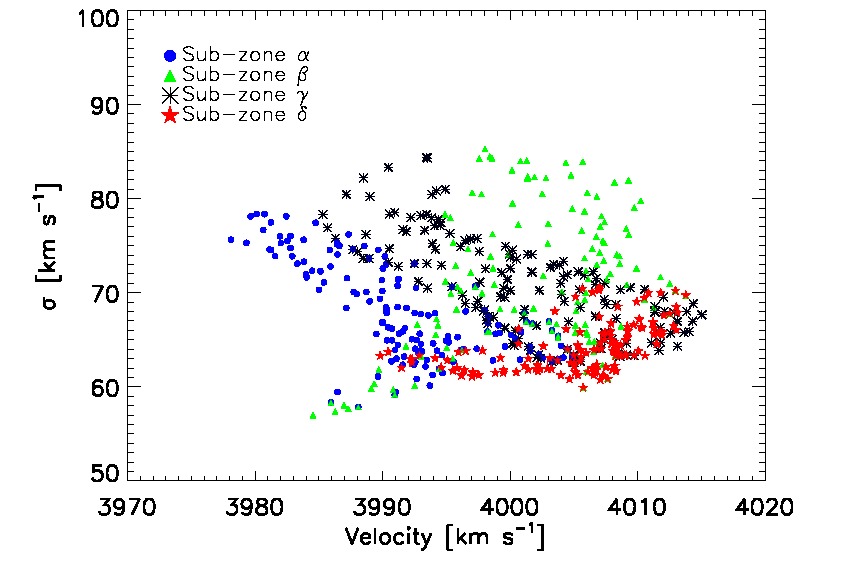}\ \includegraphics[width=\columnwidth]{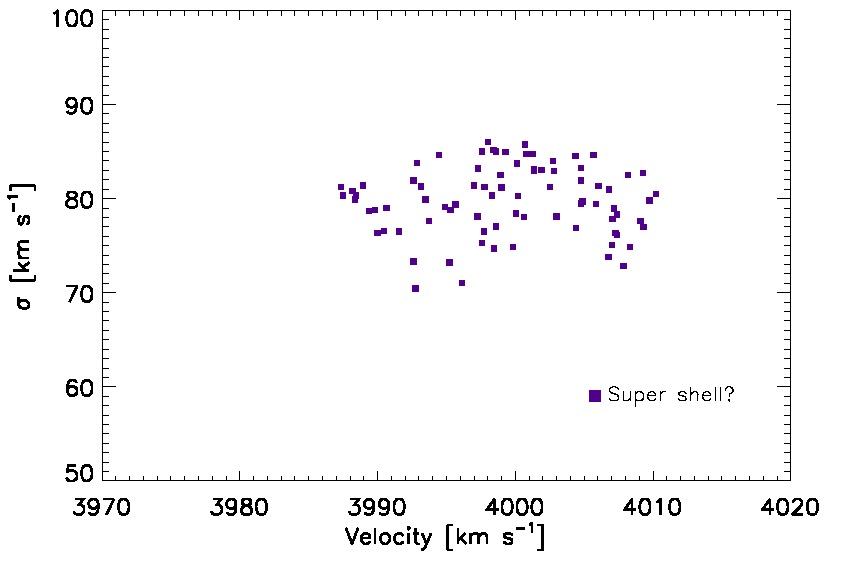}\\
\caption{\small{Top panels: $\sigma$ versus flux diagram. Middle panels: radial velocity versus flux diagram. Bottom 
panels: $\sigma$ versus radial velocity diagram. The left column diagrams include the sub-zones in Zone 4, $\alpha$, $\beta$,
$\gamma$ and $\delta$, which are represented in blue circles, green triangles, black asterisks and red stars, respectively.
The right column diagrams include the points in the area in the fourth panel in Figure \ref{mapas_diagrama} (purple square) 
represented by purple squares, which corresponds to the super shell produced by the possible super stellar cluster in sub-zone
$\beta$ (0.58''$\times$0.58'', 158$\times$158 pc).}}
\label{diagrama4}
\end{figure*}

\subsubsection{Inspecting Zone 4}

Considering the distinctive behavior of Zone 4 in the three different diagnostic diagrams with respect to the others
(which suggest a different and more complicated broadening mechanism), we have subdivided it in four sub-zones called as 
$\alpha$, $\beta$, $\gamma$ and $\delta$ (see Figure \ref{mapas_diagrama}) in order to analyze with more details the phenomena
in this zone. This division was made in an attempt to determine more accurately the number of structures present - while we expect at least two, there may be more. 

We have included these three diagnostic diagrams for the four sub-zones in the left column of Figure \ref{diagrama4}, where the
sub-zones $\alpha$, $\beta$, $\gamma$ and $\delta$ are represented by blue circles, green triangles, black asterisks and red 
stars, respectively. For clarity, the super shell placed in the sub-zone $\beta$ has been highlighted with
purple squares in the right hand panels of Figure \ref{diagrama4} (corresponding to the purple square in the last panel in
Figure \ref{mapas_diagrama}).

Analyzing the ``$\sigma$'' versus ``flux'' diagram (top left panel Figure \ref{diagrama4}) we note two independent
distributions, which are indicated with turquoise arrows. The right arrow indicates a structure placed in the sub-zone $\beta$ 
(green triangles). This presents the highest velocity dispersion values, reaching  \mbox{85\,km\,s$^{-1}$}. On the other
hand, the left arrow shows a structure which seems to be composed by the other three sub-zones of Zone 4. This last structure
presents velocity dispersions in a range of \mbox{$\sim$55-80\,km\,s$^{-1}$}. This is not considering the black asterisks of 
the sub-zone $\gamma$, which present values of \mbox{80-85\,km\,s$^{-1}$}, because these points do not belong to the burst 
associated to the galaxy HCG 31C (see fourth panel in Figure \ref{mapas_diagrama}). 
In the second panel in Figure \ref{mapas_diagrama}, we note that the sub-zone $\beta$ presents the highest stellar 
continuum emission (black contours) and lowest H$\alpha$ emission with respect to the other sub-zones, especially $\alpha$, where most of the burst is located. 
This fact suggests that in sub-zone $\beta$ there is an older stellar population than the observed in the other sub-zones, 
specially in the H$\alpha$ burst. This offset between the peak in the H$\alpha$ and stellar continuum emission may be linked 
with star formation triggered by the feedback of the super star cluster. The sub-zone $\alpha$ has the most intense H$\alpha$
emission, of up to \mbox{$\sim$5$\times$10$^{-15}$\,ergs\,s$^{-1}$\,cm$^{-2}$}, and with $\sigma$ values lower than
\mbox{80\,km\,s$^{-1}$}. 
Therefore, we can not discard the existence of more than one single stellar population (SSP) placed in Zone 4. In fact, we
report the finding of at least two stellar populations, where the oldest is located in the sub-zone $\beta$, and the other
would be in the other sub-zones, mostly in sub-zone $\alpha$, which would be triggered by the feedback of the stellar cluster.

\subsubsection{Inspecting the candidate super shell}

In order to locate the points that belong to the super shell produced by the super stellar cluster
candidate located in the sub-zone $\beta$, we have highlighted a region represented by a purple square in the last panel in
Figure \ref{mapas_diagrama}, which has been plotted independently in the three diagnostic diagrams described previously 
(right panels of Figure \ref{diagrama4}) represented by purple squares. 
Inspecting the ``$\sigma$'' versus ``flux'' diagram in the top right panel in this figure, this structure appears as a region 
delimited by velocity dispersion between $\sim$73-86\,km\,s$^{-1}$ and fluxes of 
\mbox{1.0-3.3$\times$10$^{-15}$\,ergs\,s$^{-1}$\,cm$^{-2}$}, resulting in the most intense peak in this figure.
Considering that the difference in the velocity dispersion values (at least 1-sigma$\sim$13\,km\,s$^{-1}$) between the super shell
and its neighborhood, we suggest the existence of a peak in the velocity dispersion located where the super shell is
located. 

The results found in the ``$\sigma$'' versus ``flux'' diagram for the super shell is consistent with the model proposed by 
Mu\~noz-Tu\~n\'on et al. (1996). This model describes a shell where it is possible to detect the maximum $\sigma$ value and the 
minimum surface brightness in its center. While the distance to the center is growing till the internal edge of the shell, the 
surface brightness increases and $\sigma$ decreases. This super shell would correspond to the region delimited as the possible
super shell (purple squares).
Analyzing the ``flux'' versus ``radial velocity'' diagram in the middle right panel in Figure \ref{diagrama4}, we can see
a vertical pattern in the purple squares which represent the super shell in the sub-zone $\beta$. This pattern describes a wide
range of radial velocities (3985-4010 km s$^{-1}$) in a small range of intensities 
(\mbox{1.0-2.2$\times$10$^{-15}$ ergs s$^{-1}$ cm$^{-2}$}), which may indicate a local expansion of the gas. 
Finally, in the ``radial velocity'' versus ``$\sigma$'' in the bottom right panel in Figure \ref{diagrama4}, the purple squares
distribution does not present a clear pattern. This could indicate that the expansion is isotropic, with  
no defined direction. This kind of diagram and analysis (and pattern) was also performed by Bordalo, Plana \& Telles (2009), 
who found similar results for specific areas in an H{\sc ii} region in the galaxy Zw 40.

The ages estimated for the sub-zones $\alpha$ and $\beta$ are consistent with the scenario proposed above. The estimates 
(see \S \ref{wha}) give an age of $\sim$9$\pm$2\,Myrs for the super stellar cluster in sub-zone $\beta$, and of 
$\sim$5$\pm$2\,Myrs for the H$\alpha$ emission peak in sub-zone $\alpha$, obtaining a difference in age of $\sim$4\,Myrs.

\begin{table*}
\centering
\scriptsize
\caption{Main physical properties of bursts in HCG 31A+C}
\begin{threeparttable}
        {\small
       \begin{tabular}{lccccc}
\hline
Source                 & 12+log(O/H)$^{(1)}$       & N$_{e}$$^{(2)}$$^{*}$  & EW(H$\alpha$)$^{(3)}$$^{*}$ & Age$^{(4)}$$^{*}$  & Age$^{(5)}$$^{*}$ \\
                       &                           &    cm$^{-3}$           & {\AA}                       & Myr                & Myr  \\ 
  \hline 
Burst in HCG 31C$^a$   & 8.22 $\pm$ 0.17           & 82  $\pm$ 50           & 238 $\pm$ 90                &  6 $\pm$ 2         & 5 $\pm$ 2  \\
Burst in HCG 31A$^a$   & 8.44 $\pm$ 0.16           & 228 $\pm$ 120          & 338 $\pm$ 50                &  5 $\pm$ 1         & 5 $\pm$ 1   \\
\hline
\label{tabla2}
\end{tabular} 
}
\begin{tablenotes}
\item[\emph{$^a$}]{Considering a box of 1.1''$\times$1.1'', corresponding to 298$\times$298 pc.}
\item[\emph{$^{*}$}]{Mean value, where the associated uncertainties correspond to the standard desviation.}
\item[\emph{$^{(1)}$}]{Estimated by Torres-Flores et al. (2015) using the N2 calibrator published by Marino et al. (2013).}
\item[\emph{$^{(2)}$}]{Estimated using the task {\sc temden}  in the {\sc iraf} {\sc stsdas} {\sc nebular} package.}
\item[\emph{$^{(3)}$}]{Obtained with using the {\sc idl} routine {\sc fluxer} written by Christof Iserlohe.}
\item[\emph{$^{(4)}$}]{Using the measured EW(H$\alpha$) as input to the Starburst99 code, assuming instantaneous
star formation with a metallicity of Z=0.004 and a Salpeter IMF (Leitherer et al. 1999).}
\item[\emph{$^{(5)}$}]{Using the measured EW(H$\alpha$) as input to the Starburst99 code, assuming instantaneous
star formation with a metallicity of Z=0.008 and a Salpeter IMF (Leitherer et al. 1999).}
\end{tablenotes}
\end{threeparttable}
\end{table*}

\section{Discussion}

\subsection{The main physical properties of HCG 31A+C}
\label{properties}

The Hickson Compact Group 31 has been studied by several authors, in different wavelengths, over the last years. Rubin et al. (1990),
Williams et al. (1991), Iglesias-P\'aramo \& Vilchez (1997), Vilchez \& Iglesias-P\'aramo (1998), Johnson \& Conti (2000), Verdes-Monetengro et al.
(2005), Richer et al. (2003), Amram et al. (2004, 2007), L\'opez-S\'anchez et al. (2004), Mendes de Oliveira et al. (2006) and Gallagher et
al. (2010) studied the different physical processes that are taking place in this object. Several of these results suggest that the central
region of HCG 31A+C is dominated by young and strong bursts of star-formation, which have ages lower than 7 Myrs. Rubin et al. (1990) and
L\'opez-S\'anchez et al. (2004) reported the presence of Wolf-Rayet features in the spectra of this system, which is consistent with even
younger ages. Also, Mendes de Oliveira et al. (2006) found that the galaxy HCG 31C follows the luminosity-metallicity relation (in the
K-band) defined by dwarf irregular galaxies. 

In this work we have estimated a luminosity in H$\alpha$ for the region ``center'' 
in HCG 31A+C of \mbox{log(L$_{H\alpha}$)=42.06 $\pm$ 0.09 erg s$^{-1}$}. This value is higher than the estimate done by
Iglesias-P\'aramo \& Vilchez (1997), who also studied the H$\alpha$ emission in HCG 31, obtaining a luminosity in H$\alpha$
of \mbox{log(L$_{H\alpha}$)=41.66 erg s$^{-1}$} for the system HCG 31A+C. In order to correct by internal extinction, these
authors used a colour excess of E(B-V)=0.15. L\'opez-S\'anchez et al. (2004), using  the data published by Iglesias-P\'aramo 
\& Vilchez (1997) estimate an H$\alpha$ luminosity for this system of \mbox{log(L$_{H\alpha}$)=41.78 erg s$^{-1}$} and of
\mbox{log(L$_{H\alpha}$)=41.54 erg s$^{-1}$} from their own spectroscopic data. The differences among these values 
can arise from the color excess and the extinction law used. In our case, we estimate an average 
value for the color excess, which was derived from the longslit observations of the system HCG 31A+C. This correction was 
applied to the three observed FOV. In this sense, our luminosities can be over/under estimated, given the use of an average 
value for E(B-V). Also, we need to keep in mind that the dust distribution in this system is not uniform.
In fact, Gallagher et al. (2010) observed that the dust emission is higher in the interface zone between galaxies HCG 31A and 
HCG 31C. Therefore, considering we have used the slit which cross this region for the color excess estimate, this could 
over estimate the fluxes in the rest of the fields observed. However, the main part of this study is focused in the interface
region, where the slit is located, therefore this could work at least in this context. Other facts to be considered in the 
different luminosity values are the flux calibration, which is not absolute and could result in an over estimation of the
fluxes, the area over which we have integrated the luminosities and the observational technique (IFU vs longslit vs H$\alpha$ 
imaging).

The velocity field of HCG 31A+C is perturbed. Clearly, even if the grand pattern motion displays globally a rotation movement, it
does not show the fingerprint of a simple disk in rotation, given the complex structure of the system. 
This fact is consistent with the map derived by Amram et al. (2007), which covers the entire compact group.
On the other hand, the spatial resolution of the IFU observations allows distinguishing structures in the 
central region of HCG 31A+C, which present different velocity dispersion values. This fact allowed us to detect a structure with the 
highest $\sigma$ values in the central region of HCG 31A+C, which is located in one of the ``star-forming complexes'' (complex 1) detected
by Gallagher et al. (2010, their Figure 4). These authors propose that these complexes could be formed by several stellar clusters, so
the complex 1 could include the star formation burst associated to the galaxy HCG 31C and a stellar cluster, which we suggest would be 
cleaning its environment and triggering star formation in its neighborhood.

Recently, Krabbe et al. (2014) estimated the electron density for a few H{\sc ii} regions located in interacting galaxies. These authors 
found that H{\sc ii} regions in interacting galaxies display systematically higher electron densities than those derived for H{\sc ii} 
regions belonging to isolated galaxies (taken from the literature). These authors found that the mean electron densities of star-forming
regions in interacting galaxies are in the range of \mbox{N$_{e}$ = 24--532 cm$^{-3}$}, while those obtained for isolated galaxies are in
the range of \mbox{N$_{e}$ = 40--137 cm$^{-3}$}. Our estimations of electron density for HCG 31A and HCG 31C are in the range displayed 
by the interacting sample of Krabbe et al. (2014), which is consistent with the interaction scenario for HCG 31A+C.

For the first time, we report the detection of a super stellar cluster candidate which could be an evidence of current and
triggered star formation process in HCG 31A+C. In addition, despite several studies have been developed to understand the 
compact group HCG 31, the spatial resolution of the IFU observations allowed us to carry out a detailed analysis of different 
structures in the system HCG 31A+C. With this information we estimate several physical properties in this system and we describe the star
formation bursts in this region in a more detailed way.

\subsection{HCG 31 as a pre-merger object}

Because of its complex morphology, several authors have suggested different scenarios to explain the origin of HCG 31. As 
mentioned earlier, Richer et al. (2003), using Fabry-Perot and spectroscopic data, suggest that galaxies HCG 31A and HCG 31C 
are a single entity. Oppositely , Amram et al. (2007) used high resolution Fabry-Perot data to suggest that both systems are
independent late-type galaxies, in a pre-merger phase. This last scenario is also supported by the results presented in this
paper. 

In this context, the SFRs estimated can give us evidence about the merging scenario supported by Amram et al. (2007). 
Considering the area for the SFR estimation in the region ``center'', which correspond to $\sim$2.57 kpc$^{2}$, we obtain a
star formation density of \mbox{log$\Sigma$=0.55 M$_{\odot}$ yr$^{-1}$ kpc$^{-2}$} for a 
\mbox{SFR$\sim$9.2 M$_{\odot}$ yr$^{-1}$}. Although we do not have the information about the quantity of atomic and molecular gas
for HCG 31A+C, this star formation density locate this object in the upper region of the log$\Sigma_{gas}$ versus
log$\Sigma_{SFR}$ plot presented by Daddi et al. (2010, Figure 2 in that paper). This fact suggest that HCG 31A+C is a strong 
starburst object. It will be difficult to reconcile this high star formation density with a single low mass galaxy, therefore, 
these results support the scenario on which HCG 31 is a merging object and not a single entity.

\subsection{Current and triggered star formation inside HCG 31A+C}

In \S \ref{kinematics} and \S \ref{sigmavsintensity} we suggest the existence of a super stellar cluster located 
$\sim$1'' (270\,pc) from the star formation burst associated to the galaxy HCG 31C. This super stellar cluster candidate 
presents an intense stellar continuum emission, with respect to its vicinity. Besides, this object is spatially associated with
a region with high velocity dispersions up to \mbox{$\sim$85\,km\,s$^{-1}$}. This fact suggests that the super cluster would be 
cleaning its environment through a shell in expansion which should be produced by the strong stellar winds and/or the energy 
coming from supernovae events. The strong interactions of these winds and the environment seems to be triggering star formation 
in its neighborhood, since we can detect a strong H$\alpha$ emission 270\,pc away.
In order to check if this scenario is possible, we can estimate the velocity of the expanding bubble associated with the super
stellar cluster, in order to determine if this quantity is in the range of expected values. Considering that the distance from the
stellar cluster to the H$\alpha$ peak is about $\sim$270 pc, and taking into account that the age of the complex is about 
5\,Myrs (this work and L\'opez-S\'anchez et al. 2004), we derive an expanding velocity of $\sim$53\,km\,s$^{-1}$. This value is 
consistent with the expanding velocity of some giant H{\sc ii} regions located in the Large Magellanic Cloud (LMC). For example,
Rosado et al. (1996) measured an expanding velocity of 45\,km\,s$^{-1}$ in the central region of N11. Clearly, the star-forming
complex in HCG 31 is much brighter than regions like N11, even 30 Dor, however, this estimate does not discard the scenario
proposed above. In fact, triggered star formation has been found in local galaxies, like the LMC, where it is possible to 
resolve individual stars (see review by Elmegreen 2011 on this topic). Other region widely studied is NGC 604, in the
spiral galaxy M33. 

The proposed scenario to explain the presence of the super stellar cluster candidate, which is triggering star formation in
HCG 31C, is consistent with the one proposed by Tosaki et al. (2007) for the giant H{\sc ii} region NGC 604. These authors detected a
high CO(J=3-2)/CO(J=1-0) ratio gas with an arclike distribution (called by the authors ``high-ratio gas arc'')
around the stellar cluster of NGC 604. These authors proposed a relation between the \textit{``high-ratio gas arc''},
the stellar cluster and the ``new'' stars in the H$\alpha$ shell, suggesting the following scenario:
In a first place, there would be what they called a ``first-generation star formation" associated to the central stellar cluster. Then, 
by some common phenomenon such as stellar winds and/or supernova explosions, there should be compression of the surrounding 
ISM which would result in a dense gas layer formed like an arclike distribution (the \textit{``high-ratio gas arc''}).
Then, new stars would be formed within the dense gas layer, what they called a ``second-generation star formation", which is
triggered by the stellar cluster (the first-generation stars). If this proposed scenario is correct, one could then observe this second 
star population in the mentioned dense gas layer in two ways, as shells emitting in H$\alpha$ and compact radio continuum sources.
Tosaki et al. (2007) mentioned that these facts suggest that the H{\sc ii} regions in the ``high-ratio gas'' arc are regions with current
star formation, and the regions with past star formation corresponds to the central star cluster, which has dispersed the ISM around.
The scenario proposed by Tosaki et al. (2007) for NGC 604 was supported by other authors
(e.g. Mart\'inez-Galarza et al. 2012, through the use of infrared data). In order to confirm this scenario in the central 
region of HCG 31A+C we would need new infrared and CO information.

In addition, the scenario presented in this work is consistent with the ``two-stage starburst'' model proposed by Walborn \& 
Parker (1992). In this model there is a concentrated star-forming burst which triggers a second burst after 2\,Myrs. This second
generation of stars is formed in the molecular clouds located in the periphery of the primary star-forming burst, which would
be caused by the energy coming from massive stars. 
We found a difference of 4\,Myrs between the super star cluster and the star-forming burst in HCG 31C, and considering the 
uncertainties, the scenario we suggest is consistent with this model.
This star-forming mechanism is described by Elmegreen (1998), who called it ``large scale triggering: shells or rings''. 
In this mechanism the ionized gas expansion triggers shell formation, which triggers star formation in its periphery. 
Elmegreen (1998) points out that the shell size depends on the pressure and environment density, where in low density environments
the shell can reach a size of a few hundred parsecs. These values are reached by the super shell,  which we suggest is
triggering star formation in HCG 31A+C ($\sim$300\,pc).

\section{Conclusions}

For the first time, Integral Field Spectroscopy data of the merging system HCG 31 is presented, specifically of HCG 31A+C.
With all the data analysis and the different estimates of physical properties for this system, we suggest a triggered star
formation scenario, where we propose the existence of a super stellar cluster candidate next to the star forming burst 
associated with HCG 31C. The strong stellar winds associated with this source seem to be triggering a star formation 
event in its neighborhood. Our data set supports the scenario that suggests that the central region of HCG 31 is a merger 
between two late-type spiral galaxies. This merger process has triggered the star formation in the center of HCG 31, namely,
HCG 31A+C. The study we present in this paper shows the power of IFU data in analyzing interactions in merging galaxies.

\vspace{0.5cm}

{\bf ACKNOWLEDGEMENTS We thank to the anonymous referee for his/her useful comments that have improved this paper.
MA-C acknowledges the financial support of the Direcci\'on de Investigaci\'on of the Universidad de La Serena, through a 
``Concurso de Apoyo a Tesis 2013'', under contract PT13146. ST-F acknowledges the financial support of the Chilean agency 
FONDECYT through a project ``Iniciaci\'on en la Investigaci\'on'', under contract 11121505 and the financial support of the
project CONICYT PAI/ACADEMIA 7912010004. CMdO acknowledges funding from FAPESP and CNPq. We thank James Turner for providing 
his program ``pymosaic" for the data cubes combination.}

\clearpage


\vspace{0.15cm}

\begin{thebibliography}{10}

\bibitem[\protect\citeauthoryear{ }{}]{} Allington-Smith, J., Murray, G., Content, R.Dodsworth, G., Davies, R., Miller, B. W., Jorgensen, I., Hook, I.Crampton, D., and Murowinski, R., 2002, PASP, 114, 892
\bibitem[\protect\citeauthoryear{ }{}]{} Amram, P., Plana, H., Mendes de Oliveira, C., Balkowski, C., \& Boulesteix, J. 2003, A\&A, 402, 865
\bibitem[\protect\citeauthoryear{ }{}]{} Amram, P., Mendes de Oliveira, C., Plana, H., Balkowski, C., Hernandez, O. 2007, A\&A, 471, 753
\bibitem[\protect\citeauthoryear{ }{}]{} Amram, P., Mendes de Oliveira, C., Plana, H.,Balkowski, C., Hernandez, O., Carignan, C., Cypriano, E. S.,Sodr\'e, Jr., L., Gach, J. L., and Boulesteix, J., 2004, ApJ, 612, L5
\bibitem[\protect\citeauthoryear{ }{}]{} Allende Prieto, C., Lambert, D. L., \& Asplund, M. 2001, ApJ, 556, 63
\bibitem[\protect\citeauthoryear{ }{}]{} Baldwin, J. A., Phillips, M. M., Terlevich, R. 1981, PASP, 93
\bibitem[\protect\citeauthoryear{ }{}]{} Barb\'a, R. H., Rubio, M., Roth, M. R., \& Garc\'ia, J. 2003, AJ, 125, 19
\bibitem[\protect\citeauthoryear{ }{}]{} Barb\'a, R. H., Ma\'iz Apell\'aniz, J., P\'erez, E., Rubio, M., Bolatto, A., Fari\~na, C., Bosch, G., \& Walborn, N. R. 2009, Ap\&SS, 324, 309
\bibitem[\protect\citeauthoryear{ }{}]{} Bordalo V., Plana H., Telles E., 2009, ApJ, 696, 1668
\bibitem[\protect\citeauthoryear{ }{}]{} Calzetti D., Armus L., Bohlin R. C., Kinney A. L., Koornneef J., Storchi-Bergmann T., 2000, ApJ, 533, 682
\bibitem[\protect\citeauthoryear{ }{}]{} Cedr\'es, B., Cepa, J., \& Tomita, A. 2005, ApJ, 634, 1043
\bibitem[\protect\citeauthoryear{ }{}]{} Cid Fernandes, R., Stasi\'nska, G., Schlickmann, M., Mateus, A., Asari N. V., Schoenell, W., Sodr\'e, L., 2010, MNRAS, 403, 1036.
\bibitem[\protect\citeauthoryear{ }{}]{} Daddi, E., Elbaz, D., Walter, F., Bournaud, F.,Salmi, F., Carilli, C., Dannerbauer, H., Dickinson, M., Monaco, P.,and Riechers, D., 2010, ApJ, 714, L118
\bibitem[\protect\citeauthoryear{ }{}]{} De Robertis, M.M., Dufour, R.J., \& Hunt, R.W. 1987, JRASC, 81, 195
\bibitem[\protect\citeauthoryear{ }{}]{} Dom\'inguez, A., Siana, B., Henry, A. L., Scarlata, C., Bedregal, A.G., Malkan, M., Atek, H., Ross, et al., 2013, ApJ, 763, 145
\bibitem[\protect\citeauthoryear{ }{}]{} Elmegreen, B. G. 1998, in Origins, ASP Conf. Ser. 148, ed. C. E. Woodward, J. M. Shull, \& H.A. Thronson Jr. (San Francisco, CA: ASP), 150
\bibitem[\protect\citeauthoryear{ }{}]{} Elmegreen, B. G. 2011, in EAS Publications Series, Vol. 51, Star Formation in the Local Universe, ed. C. Charbonnel \& T. Montmerle (Cambridge: Cambridge Univ. Press)
\bibitem[\protect\citeauthoryear{ }{}]{} Eufrasio, R. T., Dwek, E., Arendt, R. G., de Mello,D. F., Gadotti, D. A., Urrutia-Viscarra, F., Mendes de Oliveira, C., and Benford, D. J., 2014, ApJ, 795, 89
\bibitem[\protect\citeauthoryear{ }{}]{} Fari\~na, C., Bosch, G. L., \& Barb\'a, R. H. 2012, AJ, 143, 43
\bibitem[\protect\citeauthoryear{ }{}]{} Fiorentino, G., Clementini, G., Marconi, M.,Musella, I., Saha, A., Tosi, M., Contreras Ramos, R., Annibali, F., Aloisi, A., and van der Marel, R., 2012, 341, 143
\bibitem[\protect\citeauthoryear{ }{}]{} Fitzpatrick E. L., 1999, PASP, 111, 63
\bibitem[\protect\citeauthoryear{ }{}]{} Gallagher, S. C., Durrell, P. R., Elmegreen, D. M., Chandar, R.,English, J., Charlton, J. C., Gronwall, C., Young, J., et al., 2010, AJ, 139, 545
\bibitem[\protect\citeauthoryear{ }{}]{} Hickson, P. 1982, ApJ, 255, 382
\bibitem[\protect\citeauthoryear{ }{}]{} Hickson, P., Mendes de Oliveira, C., Huchra, J. P., and Palumbo, G. G., 1992,  ApJ, 399, 353
\bibitem[\protect\citeauthoryear{ }{}]{} Hook, I. M.; Jorgensen, I., Allington-Smith, J. R., Davies, R. L., Metcalfe, N., Murowinski, R. G., Crampton, D. 2004, PASP, 116, 425
\bibitem[\protect\citeauthoryear{ }{}]{} Iglesias-P\'aramo, J., \& V\'ilchez, J. M. 1997, ApJ, 479, 190
\bibitem[\protect\citeauthoryear{ }{}]{} Iglesias-P\'aramo, J., Buat, V., Takeuchi, T. T., Xu, K., Boissier, S., Boselli, A., Burgarella, D., Madore, B. F.,  et al., 2006, ApJS, 164, 38
\bibitem[\protect\citeauthoryear{ }{}]{} Johnson, K. E., \& Conti, P. S. 2000, AJ, 119, 2146
\bibitem[\protect\citeauthoryear{ }{}]{} Kauffmann, G., Heckman, T. M., Tremonti, C., Brinchmann, J.Charlot, S., White, S. D. M., Ridgway, S. E., Brinkmann, J., et al., 2003, MNRAS, 346, 1055
\bibitem[\protect\citeauthoryear{ }{}]{} Kennicutt, R. C., Jr. 1998, ARA\&A, 36, 189
\bibitem[\protect\citeauthoryear{ }{}]{} Kewley L. J., Dopita M. A., Sutherland R. S., Heisler C. A., Trevena J., 2001, ApJ, 556, 121 
\bibitem[\protect\citeauthoryear{ }{}]{} Krabbe, A. C., Rosa, D. A., Dors, O. L., Pastoriza, M. G., Winge, C., Hagele, G. F., Cardaci, M. V., and Rodrigues, I., 2014, MNRAS, 437, 1155
\bibitem[\protect\citeauthoryear{ }{}]{} Leitherer, C., \& Heckman, T. M. 1995, ApJS, 96, 9
\bibitem[\protect\citeauthoryear{ }{}]{} Leitherer, C., Schaerer, D., Goldader, J. D., Gonz\'alez Delgado, R. M., Robert, C., Kune, D. F., de Mello, D. F., Devost, D., and Heckman, T. M., 1999, ApJ, 123, 3
\bibitem[\protect\citeauthoryear{ }{}]{} L\'opez, L. A., Krumholz, M. R., Bolatto, A. D., Prochaska, J. X.,Ramirez-Ruiz, E., and Castro, D., 2014, ApJ, 795, 121
\bibitem[\protect\citeauthoryear{ }{}]{} L\'opez-S\'anchez, A. R., Esteban, C., Rodr\'iguez, M. 2004, ApJS, 153, 243
\bibitem[\protect\citeauthoryear{ }{}]{} Ma\'iz-Apell\'aniz, J., P\'erez, E., \& Mas-Hesse, J. M. 2004, AJ, 128, 1196
\bibitem[\protect\citeauthoryear{ }{}]{} Marino, R. A., Rosales-Ortega, F. F., S\'anchez, S. F., Gil de Paz,A., V\'ilchez, J., Miralles-Caballero, D., Kehrig, C., P\'erez-Montero,E.,  et al., 2013, AAP, 559, A114
\bibitem[\protect\citeauthoryear{ }{}]{} Mart\'inez-Delgado I., Tenorio-Tagle G., Mu\~noz-Tu\~n\'on C., Moiseev A. V., Cair\'os L. M., 2007, AJ, 133, 2892
\bibitem[\protect\citeauthoryear{ }{}]{} Mart\'inez-Galarza J. R., Hunter D., Groves B., Brandl B., 2012, ApJ, 761, 3
\bibitem[\protect\citeauthoryear{ }{}]{} Mendes de Oliveira, C. L., Temporin, S., Cypriano, E. S., Plana, H., Amram, P., Sodr\'e, L. J., \& Balkowski, C. 2006, AJ, 132, 570
\bibitem[\protect\citeauthoryear{ }{}]{} Moiseev A. V., \& Lozinskaya T. A., 2012, MNRAS, 423, 1831
\bibitem[\protect\citeauthoryear{ }{}]{} Mu\~noz-Tu\~n\'on C., Tenorio-Tagle G., Casta\~neda H. O., Terlevich R., 1996, AJ, 112, 1636
\bibitem[\protect\citeauthoryear{ }{}]{} Osterbrock D. E., 1989, Astrophysics of Gaseous Nebulae and Active Galactic Nuclei. University Science Books, Mill Valley, CA
\bibitem[\protect\citeauthoryear{ }{}]{} Perez J., Michel-Dansac L., Tissera P. B., 2011, MNRAS, 417, 580
\bibitem[\protect\citeauthoryear{ }{}]{} Plana, H., Amram, P., Mendes de Oliveira, C., Balkowski, C., \& Boulesteix, J. 2003, AJ, 125, 1736
\bibitem[\protect\citeauthoryear{ }{}]{} Richer, M. G., Georgiev, L., Rosado, M., Bullejos, A., Valdez-Guti\'errez, M., \& Dultzin-Hacyan, D. 2003, A\&A, 397, 99
\bibitem[\protect\citeauthoryear{ }{}]{} Rosado, M., Laval, A., Le Coarer, E., Georgelin, Y. P., Amram, P.,Marcelin, M., Goldes, G., and Gach, J. L., 1996, A\&A, 308, 588
\bibitem[\protect\citeauthoryear{ }{}]{} Rubin, V. C., Hunter, D. A., \& Ford, W. K. J. 1990, ApJ, 365, 86
\bibitem[\protect\citeauthoryear{ }{}]{} Rupke, D. S. N., Kewley, L. J., Barnes, J. E. 2010a, ApJ, 710L, 156
\bibitem[\protect\citeauthoryear{ }{}]{} S\'anchez, S. F., Kennicutt, R. C., Gil de Paz, A., van de Ven, G.,Vílchez, J. M., Wisotzki, L., Walcher, C. J., Mast, D., et al., 2012, A\&A, 538, A8
\bibitem[\protect\citeauthoryear{ }{}]{} Stasi\'nska G., Cid Fernandes R., Mateus A., Sodr\'e L., Asari N. V., 2006, MNRAS, 371, 972
\bibitem[\protect\citeauthoryear{ }{}]{} Torres-Flores, S., Mendes de Oliveira, C., Amram, P., Alfaro-Cuello, M., Carrasco, E. R. \& de Mello, D. F., 2015, ApJ, 798, L24
\bibitem[\protect\citeauthoryear{ }{}]{} Tosaki, T., Miura, R., Sawada, T., Kuno, N., Nakanishi, K., Kohno, K., Okumura, S. K., and Kawabe, R., 2007, ApJ, 664, L27
\bibitem[\protect\citeauthoryear{ }{}]{} Verdes-Montenegro, L., Del Olmo, A., Yun, M. S., \& Perea, J. 2005, A\&A, 430, 443
\bibitem[\protect\citeauthoryear{ }{}]{} Walborn, N. R., \& Parker, J. W. 1992, ApJ, 399, L87
\bibitem[\protect\citeauthoryear{ }{}]{} Williams, B. A., McMahon, P. M., \& van Gorkom, J. H. 1991, AJ, 101, 1957
\bibitem[\protect\citeauthoryear{ }{}]{} V\'ilchez, J. M \& Iglesias-P\'aramo, J. 1998, ApJ, 508, 248
\end{thebibliography}
\end{document}